\documentclass[a4paper,11pt]{article}
\pdfoutput=1 

\usepackage{jcappub} 

\usepackage[T1]{fontenc} 
\usepackage{graphicx}
\usepackage{subcaption}
\usepackage{url}

\def\fnl{f_{\mathrm{NL}}}

\title{\boldmath Efficient and Optimal Analysis of Galaxy Surveys on Large Scales }

\author[a, b]{Julius Adolff}
\author[a, b]{and Uroš Seljak}
\affiliation[a]{Department of Physics, University of California, Berkeley, CA 94720, USA}
\affiliation[b]{Lawrence Berkeley National Laboratory, University of California, Berkeley, CA 94720, USA}

\emailAdd{juliusadolff@berkeley.edu}

\abstract{On sufficiently large scales the galaxy overdensity field is approximately Gaussian, so its information can be losslessly compressed into the power spectrum. Standard power spectrum estimators are nevertheless suboptimal on the largest scales, where the survey window induces significant coupling between different modes. Optimal power spectrum analysis typically requires an inverse of the covariance matrix, which can be overly expensive. Here we develop an alternative field level approach, which is significantly faster. We assess its impact on constraining primordial non-Gaussianity of the local type, which is particularly sensitive to these largest scales. Relative to the standard power spectrum analysis, our field level likelihood forecasts a $14\%$ improvement in constraining power. For the DESI DR2 LRG and QSO samples, our approach forecasts $\sigma(f_{\rm NL})=4.1$, compared to $\sigma(f_{\rm NL})=4.8$ from the power spectrum based analysis. The field level approach also enables efficient implementations of additional non-cosmological contributions to the signal. As an example, we present a novel non-parametric method to mitigate angular systematics, enabling efficient deprojection of contaminated modes.
}

\begin{document}
\maketitle
\flushbottom

\section{Introduction}
\label{sec:intro}
Clustering of galaxies in redshift space contains a wealth of cosmological information~\cite{Desjacques_2018, Ivanov_2020, Adame_2025}. Extracting it requires an accurate statistical model of the galaxy overdensity field and a tractable method for evaluating the resulting likelihood. This field level analysis is particularly challenging when much of the information lies on nonlinear scales, where the forward model and the likelihood remain active areas of research \cite{Cabass_2020, parker2026fieldlevelbaryonacousticoscillation, Kosti_2023, leclercq2015probabilisticcartographylargescalestructure}. Primordial non-Gaussianity offers a comparatively simpler setting. Its strongest clustering signature is on very 
large scales, which are approximately linear \cite{Dalal_2008, Desjacques_2010}, and both the field and the likelihood 
can be modeled more reliably.

In this regime the central challenge is to extract the available information efficiently while accounting for observational effects, including integral constraints \cite{de_Mattia_2019}, the survey window \cite{Yamamoto_2003, sailer2026optimalexactwideanglepower}, and imaging systematics \cite{chaussidon2025constrainingprimordialnongaussianitydesi}. On linear scales, the field level likelihood is Gaussian, but evaluating it directly is computationally impractical. A complete extraction of information from a modern survey like DESI requires voxelization of the survey into $10^6$ or more linear modes \cite{desicollaboration2016desiexperimentisciencetargeting}, and survey geometry couples these modes, producing a dense covariance matrix. In an optimal analysis this 
matrix needs to be inverted, making 
this approach extremely expensive. 
Analyses therefore typically compress the data into a summary power spectrum or correlation function statistic and approximate their likelihood as Gaussian, with a covariance estimated from mocks at fiducial cosmological and bias parameters \cite{Li_2019, chaussidon2025constrainingprimordialnongaussianitydesi}. This Gaussian 
approximation is however 
not the true likelihood, and we will show that, for parameters constrained by relatively few modes such as $\fnl$, a fixed covariance approximation can lead to miscalibrated posteriors.

A standard summary statistic is the power spectrum, estimated using weights that depend only on redshift \cite{Feldman_1994}. Effects such as the survey window are then theoretically modeled and calibrated using mocks \cite{chaussidon2025constrainingprimordialnongaussianitydesi}. This approach can be suboptimal on the largest scales, where the window induces appreciable mode coupling. Building on previous work \cite{Tegmark_1998, Seljak_1998, Horowitz_2019, Philcox_2021, Hamilton_1997, Millea_2022}, we instead construct field level quadratic estimators for the parameters of interest. These estimators are locally optimal under the Gaussian model and incorporate observational effects, such as the survey window, through the forward model. When the forward model is linear in the unknown parameters, we can characterize the parameter dependence of the compressed statistic exactly, without assuming a parameter independent covariance. 

Working at the field level also allows us to account for observational effects {before} compressing the data. We model the survey window directly and project out modes affected by uncertain radial or angular selection functions. 
Known angular systematics can be represented by templates and marginalized over through a projection \cite{Weaverdyck_2021}. We also consider the more challenging case in which the angular systematic is not described by a known template. In this case, we treat the systematic field as an unknown component and marginalize over it at the field level, without imposing an arbitrary scale cut and with minimal loss of cosmological information.

This paper is organized as follows. We first describe the field level forward model and our treatment of the survey window and integral constraints. We then derive the optimal quadratic summary statistic, motivated by a local Gaussian approximation to the field level likelihood, and its statistical description. Finally, we forecast constraints on $\fnl$ from the DESI DR2 LRG and QSO samples and compare them with those from a standard power spectrum analysis. We conclude by discussing the implications and limitations of the approach.

\section{Theoretical Model}\label{sec:Theory1}
The primary observable in a galaxy survey is the density field \cite{Feldman_1994},
\begin{equation}
    d(\boldsymbol x)=n_g(\boldsymbol x)-\alpha\,n_r(\boldsymbol x),
\end{equation}
where the weighted galaxy and random catalog number density fields are, respectively,
\begin{equation}
    n_g(\boldsymbol x)=\sum_{g}w_g\,\delta_D(\boldsymbol x-\boldsymbol x_g),
    \qquad
    n_r(\boldsymbol x)=\sum_{r}w_r\,\delta_D(\boldsymbol x-\boldsymbol x_r).
\end{equation}
The weights account for effects such as completeness, imaging systematics,\footnote{We will, however, discuss an alternative treatment of imaging systematics below and therefore do not include $w_{\rm sys}$ in our analysis.} and spectroscopic efficiency \cite{ross2024constructionlargescalestructurecatalogs}. The random catalog traces the survey selection function, including its geometry, and can be constructed using methods such as redshift shuffling to impose the appropriate integral constraints \cite{Ross_2012}. We define the expected weighted number density as
\begin{equation}
    \bar n(\boldsymbol x)\equiv\langle n_g(\boldsymbol x)\rangle
    =\alpha\,\langle n_r(\boldsymbol x)\rangle.
\end{equation}
It varies across the survey because of the survey window and the redshift evolution of the tracer population.

On sufficiently large scales, we can model $\boldsymbol d$ as a Gaussian field \cite{Bernardeau_2002} sourced by the Gaussian matter field $\boldsymbol \delta$ and Gaussian noise $\boldsymbol n$. Specifically, we write
\begin{equation}
    \boldsymbol d = \boldsymbol f(\boldsymbol \delta, \boldsymbol \theta) + \boldsymbol n, \label{eq:DataModel}
\end{equation}
where $\boldsymbol f$ is the \textit{forward model}, described below, and depends on the cosmological parameters $\boldsymbol \theta$. The fields $\boldsymbol \delta$ and $\boldsymbol n$ are Gaussian, with known covariances \cite{Feldman_1994, Bernardeau_2002}
\begin{align}
    \langle n(\boldsymbol x)n(\boldsymbol y)^*\rangle &= \delta_D(\boldsymbol x - \boldsymbol y)\,\bar n(\boldsymbol x)\,(\langle w_g^2 \rangle + \alpha \langle w_r ^2\rangle ), \\
    \langle \delta (\boldsymbol k)\delta(\boldsymbol q)^*\rangle &= (2\pi)^3\delta_D(\boldsymbol k - \boldsymbol q)\,P(k).
\end{align}
In the absence of a nontrivial window or redshift evolution of the tracer, the noise covariance is proportional to the identity, and the data covariance $\mathbf C = \langle \boldsymbol d \boldsymbol d^\dagger \rangle $ is diagonal. In this case, the data $\boldsymbol d$ can be summarized by its two-point function, which reduces to a one dimensional function (the diagonal of $\mathbf{C}$) and provides a lossless summary statistic. The survey window and redshift evolution complicate this picture by coupling different Fourier modes in the data \cite{Beutler_2019}. These effects are usually captured by a window matrix that relates $P(k)$ to the observed two-point function. When estimating the two-point function, optimal galaxy weighting is typically approximated by taking the diagonal limit of $\mathbf C$ in real space \cite{Feldman_1994}.

Here, we propose an alternative method that treats the observed data $\boldsymbol{d}$ at the field level, allowing information to be extracted optimally. Building on the methods of \cite{Horowitz_2019, Seljak_2017, Seljak_1998, Millea_2022}, we extend the Marginal Unbiased Score Expansion (MUSE) framework. Our forward model, which is linear in both the matter field $\boldsymbol \delta$ and the parameters $\boldsymbol \theta$ (a property that will be important later), has three components:
\begin{equation}
    \boldsymbol f = \mathbf \Pi \mathbf W\mathbf R = \mathbf \Pi_{\rm AIC}\mathbf \Pi_\Lambda\mathbf \Pi_{\rm RIC}\mathbf W\mathbf R,
\end{equation}
where $\mathbf R$ describes bias and redshift space distortions (RSD), $\mathbf W$ is the window operator, and $\mathbf \Pi$ projects onto the subspace of the data that we can model accurately. This subspace is limited by our restriction to linear scales, represented by $\mathbf \Pi_\Lambda$, and by incomplete knowledge of the selection function, represented by the angular and radial integral constraint projectors $\mathbf \Pi_{\rm AIC}$ and $\mathbf \Pi_{\rm RIC}$. We discuss these components in turn below.

Before comparing the model with the data, we restrict the data to the modes accessible to our analysis by making the substitution $\boldsymbol d \rightarrow  \boldsymbol{d}_\Pi = \mathbf \Pi\boldsymbol d$. Because the data contain both signal and noise, this projection also changes the noise covariance. Defining $\mathbf N = \langle \boldsymbol n \boldsymbol n^\dagger \rangle $, we obtain $\mathbf N \rightarrow \mathbf N_\Pi = \mathbf \Pi\, \mathbf N\,\mathbf \Pi^\dagger$. Finally, with $\mathbf P = \langle \boldsymbol \delta \boldsymbol \delta^\dagger \rangle $, the full covariance of the projected data is
\begin{equation}
    \mathbf C_\Pi = \mathbf \Pi \langle \boldsymbol d\boldsymbol d^\dagger \rangle \mathbf \Pi^\dagger = \boldsymbol f\, \mathbf P\,\boldsymbol f^\dagger + \mathbf N_\Pi. 
\end{equation}
This covariance is singular because the projection removes modes from the data. We address this by inverting $\mathbf C_\Pi$ only on the subspace onto which $\mathbf \Pi$ projects, as occurs automatically when we use the conjugate gradient method. We discuss this procedure in more detail in subsequent sections.

\subsection{Bias, Redshift Space Distortions and the Window}\label{sec:Window}
The \textit{bias expansion} $\mathbf R$ is modeled as a linear function of the parameters $\boldsymbol \theta = (b_1, \Sigma_s, f_{\rm NL})$. Placing the observer at $\boldsymbol o$ and letting $\partial_r$ denote the derivative in the radial direction from the observer, we relate the galaxy and matter fields through \cite{Castorina_2018, Desjacques_2018, Seljak_2011, Jackson_1972}
\begin{equation}
    \mathbf R \boldsymbol \delta = D(z)\left(\mathbf b + f(z) \frac{\partial^2 }{\partial r^2}\nabla^{-2} + \Sigma_s \frac{\partial ^2}{\partial r^2} \right) \boldsymbol\delta ,
\end{equation}
where the bias operator $\mathbf b$ is a 
complicated non-local expression in real space but acts as a  multiplication in Fourier space, and includes the scale dependent bias introduced by the primordial signal~\cite{Dalal_2008, Slosar_2008}
\begin{equation}
    \mathbf{b} = b_1(z) +  \frac{\fnl b_\phi(z)}{D(z)\,\mathcal M(k)}.
\end{equation}
We model the redshift evolution of the linear bias as \cite{Laurent_2017, chaussidon2025constrainingprimordialnongaussianitydesi}
\begin{equation}
    b_1(z) = b_1 +a\,{(1 + z)^2} \label{eq:BiasEvolution}
\end{equation}
and set $b_\phi(z) = 2 \delta_c\,(b_1(z) - p)$, where $p$ depends on the tracer. We assume that $a$ is known from an analysis of small scale clustering, similar to the treatment in the DESI analysis \cite{chaussidon2025constrainingprimordialnongaussianitydesi}.\footnote{A more conservative analysis may also allow $a$ to vary; however, this would complicate a direct comparison with the DESI results.} We evaluate the derivatives in Fourier space using ${\hat{\boldsymbol r} = (\boldsymbol{x} - \boldsymbol{o})/|\boldsymbol{x} - \boldsymbol{o}|}$ and the substitution
\begin{equation}
    \frac{\partial^2}{\partial r ^2} \rightarrow -\sum_{ij}\hat r_i\hat r_j{k_ik_j}.
\end{equation}
Treating RSD in this manner avoids the need for wide angle corrections. Specifically, we define the tensors
\begin{equation}
    \beta_{ij} = \mathtt{iFFT}\left[ \frac{k_i k_j}{k^2}\,\delta(k) \right], \qquad \gamma_{ij} = -\mathtt{iFFT}\left[ {k_i k_j}\,\delta(k) \right],
\end{equation}
along with the scalar $\phi = \mathtt{iFFT}[\delta(k)/\mathcal M(k)]$, to obtain
\begin{align*}
    (\mathbf R\boldsymbol \delta)(\boldsymbol x) &= D(z)b_1(z)\,\delta(\boldsymbol x) + \fnl b_\phi(z)\,\phi(\boldsymbol{x}) \\ &\quad+ D(z)\,\sum_{ij} \hat r_i\hat r_j\left[f(z)\beta_{ij}(\boldsymbol{x}) + \Sigma_s \gamma_{ij}(\boldsymbol{x})  \right].
\end{align*}
We apply the {window} to $\mathbf R\boldsymbol\delta$ by multiplying with the expected number density $\bar n(\boldsymbol x)$
\begin{equation}
    (\mathbf R\boldsymbol \delta)(\boldsymbol x)  \rightarrow \bar n(\boldsymbol x) (\mathbf R\boldsymbol \delta)(\boldsymbol x).
\end{equation}
To avoid aliasing, we perform this multiplication on a padded mesh twice the size of the original mesh. This ensures that the real space product exactly represents the relevant Fourier space convolution. We denote by $\mathbf W$ the linear operator that pads the field, multiplies it by $\bar n(\boldsymbol x)$ in real space, and then reduces it back to the original Fourier space mesh. We write
\begin{equation}
    \boldsymbol{\Delta_g} = \mathbf W\mathbf R\boldsymbol \delta. 
\end{equation}
Although $\mathbf N$ and $\mathbf W$ are proportional, they have different interpretations, so we distinguish them notationally.

\subsection{Mode Deprojection}\label{sec:IC}
We cannot model every mode in the observed data $\boldsymbol d$. Our analysis is restricted to scales on which linear theory is valid, and our description of the survey selection function is incomplete. The latter limitation gives rise to radial and angular integral constraints. We therefore apply a mode deprojection operator $\mathbf \Pi$ to both the data and the model $\boldsymbol{\Delta_g}$. The simplest component is $\mathbf \Pi_\Lambda$, which restricts the analysis to modes described by linear theory: $k \leq k_{\max} = 0.08\,h/\mathrm{Mpc}$. Because $[\mathbf W, \mathbf \Pi_\Lambda] \neq 0$, we must apply the window $\mathbf W$ \textit{before} restricting to the observational band, consistent with how the observed data are formed.

We decompose the projection into three components:
\begin{equation}
    \mathbf \Pi = \mathbf \Pi_{\rm AIC}\,\mathbf \Pi_\Lambda\mathbf \Pi_{\rm RIC}.
\end{equation}
The \textit{radial integral constraint} (RIC) $\mathbf \Pi_{\rm RIC}$ and the \textit{angular integral constraint} (AIC) $\mathbf \Pi_{\rm AIC}$ arise from uncertainty in the radial and angular selection functions and the finite size of the survey \cite{de_Mattia_2019, 1991MNRAS, 2017MNRAS, Weaverdyck_2021}. Because the \textit{true} redshift evolution of the tracer is unknown, we match the redshift distribution of the randoms to that of the observed galaxies. The radial modes removed by this procedure therefore cannot be used to extract cosmological information. Similarly, our incomplete understanding of imaging systematics, such as dust, prevents us from reliably interpreting modes correlated with the corresponding templates.

We implement the radial integral constraint by subtracting the mean in each redshift shell from the data. Writing $\boldsymbol{r}_i = \mathbf 1_{[z_i, z_i + \Delta z_i]}$ for a constant amplitude in shell $i$, we obtain
\begin{equation}
    \mathbf{\Pi}_{\rm RIC} = \mathbf 1 - \mathbf W\,\sum_i \frac{\boldsymbol{r}_i \boldsymbol{r}_i^\dagger }{ \boldsymbol{r}_i^\dagger\mathbf W\boldsymbol{r}_i},
\end{equation}
which is a projector. In practice, we use as many redshift shells as the mesh can resolve.

To treat the AIC, we introduce templates $\boldsymbol{t}_i$, where the index $i$ runs over both the templates and the redshift bins in which their amplitudes are fitted independently. We allow these amplitudes to vary freely:
\begin{equation}
    \boldsymbol{\Delta_g} \rightarrow \boldsymbol{\Delta_g} + \sum_i a_i\, \mathbf W\boldsymbol{t}_i.
\end{equation}
Assigning a flat prior to each template amplitude and marginalizing over these amplitudes is equivalent to discarding the subspace spanned by the templates. We take the projected templates $\{ \mathbf \Pi_\Lambda\mathbf \Pi_{\rm RIC}\mathbf W\boldsymbol{t}_i\}$ as a starting basis and, using Gram–Schmidt orthogonalization, construct a basis orthonormal with respect to $\mathbf N^{-1}$. This choice weights each voxel according to its Poisson noise. Denoting the resulting basis by $\{\boldsymbol e_i\}$, we obtain the corresponding amplitudes through matched filtering:
\begin{equation}
    \hat{a}_i = \boldsymbol e_i^\dagger \mathbf N^{-1}\boldsymbol d.
\end{equation}
We may then write the resulting estimate as
\begin{equation}
    \boldsymbol{\Delta_g} = \left(\boldsymbol 1 - \sum_i {\boldsymbol e_i\boldsymbol e_i^\dagger \mathbf N^{-1}} \right)\boldsymbol d.
\end{equation}
Thus, fitting the free amplitudes $a_i$ is equivalent to applying the projector
\begin{equation}
    \mathbf \Pi_{\rm AIC} = \boldsymbol 1 - \sum_i {\boldsymbol e_i\boldsymbol e_i^\dagger \mathbf N^{-1}}.
\end{equation}
Applying this projector to both the model and the data removes the need to fit the amplitudes explicitly and retains the subspace orthogonal to the templates. Applying $\mathbf \Pi_{\rm AIC}$ to a vector $\boldsymbol{v}$ is inexpensive because $\mathbf N$ is diagonal in real space. In particular,
\begin{equation}
    \boldsymbol e_i^\dagger \mathbf N^{-1}\,\boldsymbol{v} = \frac{1}{\langle w_g^2 \rangle + \alpha \langle w_r ^2\rangle}\sum_{\boldsymbol{x}\,\in\,\mathrm{window}}\frac{e_i^*(\boldsymbol{x})\,v(\boldsymbol{x})}{\bar n(\boldsymbol{x})},
\end{equation}
so applying $\mathbf \Pi_{\rm AIC}$ requires only a small number of inner products.

\section{Parameter Inference}\label{sec:MUSE}
The parameter likelihood $p(\boldsymbol d_\Pi\,|\,\boldsymbol \theta)$ is obtained by marginalizing over the latent field $\boldsymbol \delta$ \cite{Seljak_1998}
\begin{equation}
    p(\boldsymbol d_\Pi\,|\,\boldsymbol \theta) = \int d\boldsymbol \delta \,p(\boldsymbol d_\Pi\,|\,\boldsymbol \theta, \boldsymbol \delta)p(\boldsymbol \delta ), \label{eq:LatentIntegral}
\end{equation}
where $p(\boldsymbol d_\Pi\,|\,\boldsymbol \theta, \boldsymbol \delta)$ and $p(\boldsymbol \delta )$ are Gaussian distributions. In more general settings, evaluating the integral~\eqref{eq:LatentIntegral} is very difficult and can only be accomplished approximately by sampling the joint distribution $p(\boldsymbol d_\Pi\,|\,\boldsymbol \theta, \boldsymbol \delta)p(\boldsymbol \delta )$ before marginalizing over $\boldsymbol \delta$ \cite{Simon_2025, Schmidt_2019, Cabass_2020, Jasche_2013, parker2026fieldlevelbaryonacousticoscillation}. In the Gaussian case considered here, however, the integral can be evaluated exactly. The log-likelihood $\ell(\boldsymbol{\theta}) \equiv \log p(\boldsymbol d_\Pi\,|\,\boldsymbol \theta)$ is then \cite{Seljak_1998, Horowitz_2019}
\begin{equation}
     \ell(\boldsymbol{\theta}) = -\frac{1}{2}\boldsymbol{d}_\Pi^\dagger\,\mathbf C_\Pi(\boldsymbol{\theta})^{-1}\boldsymbol{d}_\Pi -\frac{1}{2}\log \det [(2 \pi)\mathbf C_\Pi(\boldsymbol{\theta})]. \label{eq:Likelihood}
\end{equation}
Because $\mathbf C_\Pi$ is dense, computing $\mathbf{C}_\Pi^{-1}$ and, especially, $\log \det \mathbf C_\Pi$ is prohibitively expensive. Directly sampling the likelihood is therefore impractical. Instead, we introduce a summary statistic motivated by a quadratic expansion of the likelihood around a fiducial model. We show that, within this expansion, all information about the parameters $\boldsymbol{\theta}$ is captured by $\dim \boldsymbol{\theta}$ numbers, $E_a$. In Sec.~\ref{sec:SumStat}, we derive a non-Gaussian likelihood for this summary statistic.

\subsection{Laplace Approximation}\label{sec:Laplace}

Because the forward model $\boldsymbol f$ is linear in the parameters $\boldsymbol \theta$, the posterior of $\boldsymbol \theta$ is well approximated as Gaussian. We expand the log-likelihood locally to quadratic order around a fiducial parameter estimate $\boldsymbol{\theta}_{\rm fid}$ to derive the corresponding Gaussian approximation. This estimate could, for example, come from a conventional FKP analysis. The resulting approximation is often called a {Laplace approximation}. In the present context we build on several methods previously developed in \cite{Millea_2022, Seljak_2017, Horowitz_2019}, which are presented here again only for completeness, and therefore we refer to our approach as MUSE.

We denote the gradient by $\boldsymbol{j} = \partial \ell /\partial \boldsymbol{\theta}|_{\boldsymbol{\theta}_{\rm fid}}$ and the negative Hessian (which informs the posterior covariance) by $F_{ab} = -\partial ^2\ell / \partial \theta_a\partial \theta_b|_{\boldsymbol{\theta}_{\rm fid}}$. The Newton update to $\boldsymbol{\theta}_{\rm fid}$ then satisfies
\begin{equation}
    \mathbf F\Delta \hat{\boldsymbol \theta} = \boldsymbol{j}. \label{eq:NewtonStep}
\end{equation}
From Eq.~\eqref{eq:Likelihood}, the gradient has two contributions:
\begin{equation}
    -2\,j_a = \frac{\partial}{\partial \theta_a}\boldsymbol{d}_\Pi^\dagger\,\mathbf C_\Pi(\boldsymbol{\theta})^{-1}\boldsymbol{d}_\Pi +  \frac{\partial}{\partial \theta_a}\log \det \mathbf C_\Pi. \label{eq:gradient}
\end{equation}
Evaluating the first contribution requires $\mathbf C_\Pi(\boldsymbol{\theta})^{-1}\boldsymbol{d}_\Pi$. We obtain this quantity by using a suitably preconditioned conjugate gradient method to solve
\begin{equation}
    \mathbf C_\Pi\, \boldsymbol{w} = \boldsymbol{d}_\Pi \label{eq:CGsol}
\end{equation}
iteratively, typically converging after $\sim 50$ iterations.\footnote{In practice, we also regularize $\mathbf C_\Pi$ to obtain fast convergence and therefore invert $\mathbf C_\Pi + \epsilon\mathbf \Pi$, with $\epsilon \ll1$. To test the impact of this effect, we expand $(\mathbf C_\Pi + \epsilon\mathbf \Pi)^{-1} = \mathbf C_\Pi^{-1} - \epsilon(\mathbf C_\Pi + \epsilon\mathbf \Pi)^{-1}\mathbf \Pi(\mathbf C_\Pi + \epsilon\mathbf \Pi)^{-1} + \mathcal O(\epsilon^2)$ and note that we can then write $\mathbf C_\Pi^{-1}\boldsymbol v   = (\mathbf C_\Pi + \epsilon\mathbf \Pi)^{-1}\boldsymbol v + \epsilon(\mathbf C_\Pi + \epsilon\mathbf \Pi)^{-1}\mathbf \Pi(\mathbf C_\Pi + \epsilon\mathbf \Pi)^{-1}\boldsymbol{v} + \mathcal O(\epsilon^2)$. This procedure delivers quadratic convergence, but has no discernible impact on the parameter posteriors. }

The score $j_a$ consists of a realization dependent quadratic term and a trace term. The former measures the response of the observed realization to the parameter, while the latter subtracts its ensemble mean. The quadratic term is
\begin{align}
    E_a &= -\frac{1}{2}\frac{\partial}{\partial \theta_a}\boldsymbol{d}_\Pi^\dagger\,\mathbf C_\Pi(\boldsymbol{\theta})^{-1}\boldsymbol{d}_\Pi \nonumber \\
    &= \frac{1}{2}\boldsymbol{d}_\Pi^\dagger \mathbf C_\Pi^{-1}\,\frac{\partial \mathbf C_\Pi}{\partial \theta_a}\mathbf C_\Pi^{-1}\boldsymbol{d}_\Pi \\
    &= \frac{1}{2}\boldsymbol{w}^\dagger \frac{\partial \mathbf C_\Pi}{\partial \theta_a} \boldsymbol{w}. \nonumber
\end{align}
We can write $E_a$ in a more interpretable form. Using the identity $\partial \mathbf C_\Pi/\partial \theta_a = \left({\partial \boldsymbol{f}}/{\partial \theta_a} \right)\mathbf P\boldsymbol{f}^\dagger + \boldsymbol{f}\,\mathbf P\left({\partial \boldsymbol{f}}/{\partial \theta_a} \right)^\dagger$ and defining $\hat{\boldsymbol{ \delta}} = \mathbf P\boldsymbol{f}^\dagger \mathbf C_\Pi(\boldsymbol{\theta})^{-1}\boldsymbol{d}_\Pi$ as the minimum variance estimate of the matter field given the data $\boldsymbol{d}_\Pi$, we then find by matrix manipulation\footnote{We must show that $\mathbf N_\Pi^{-1}(\boldsymbol{d}_\Pi - \boldsymbol{f}(\hat{\boldsymbol{\delta}})) = \mathbf N_\Pi^{-1}(\mathbf 1 - \boldsymbol{f}\,\mathbf P \boldsymbol{f}^\dagger \mathbf C_\Pi^{-1})\boldsymbol{d}_\Pi$ equals $\mathbf{C}_\Pi^{-1}\boldsymbol{d}_\Pi$. For this, it is sufficient to note that \begin{equation}
    \mathbf N_\Pi^{-1}(\mathbf 1 - \boldsymbol{f}\,\mathbf P \boldsymbol{f}^\dagger \mathbf C_\Pi^{-1})\mathbf C_\Pi =  \mathbf N_\Pi^{-1}\mathbf C_\Pi- \mathbf N_\Pi^{-1}\boldsymbol{f}\,\mathbf P \boldsymbol{f}^\dagger, \nonumber
\end{equation}
which, using $\mathbf N_\Pi^{-1}\mathbf C_\Pi = \mathbf N_\Pi^{-1}(\boldsymbol{f}\,\mathbf P \boldsymbol f^\dagger + \mathbf N_\Pi) = \mathbf N_\Pi^{-1}\boldsymbol{f}\,\mathbf P \boldsymbol f^\dagger + \mathbf 1$, gives the desired result.}
\begin{equation}
    E_a(\boldsymbol{d}, \boldsymbol{\theta}_{\rm fid}) = \biggl( \frac{\partial \boldsymbol f(\hat{\boldsymbol{ \delta}})}{\partial \theta_a}\biggr)^\dagger \mathbf N_\Pi^{-1}(\boldsymbol{d}_\Pi - \boldsymbol{f}(\hat{\boldsymbol{\delta}})). \label{eq:MUSEEstimator}
\end{equation}
Thus, $E_a$ compares the data with the forward model, weighting their difference by the inverse noise covariance and the model's linear response to $\theta_a$.

The second contribution to Eq.~\eqref{eq:gradient} is
\begin{equation}
    b_a(\boldsymbol{\theta}_{\rm fid}) = \frac{1}{2}\frac{\partial }{\partial \theta_a}\log \det \mathbf C_\Pi = \frac{1}{2}\mathrm{tr}\left( \mathbf C_\Pi^{-1} \frac{\partial \mathbf C_\Pi}{\partial \theta_a} \right)
\end{equation}
and is commonly called the \textit{bias} contribution \cite{Seljak_1998}. It accounts for the difference between the minimum variance estimate $\hat{\boldsymbol \delta}$ and the true matter field induced by Wiener filtering. Computing this term directly would require the trace of a high dimensional dense matrix and is therefore infeasible. Instead, we estimate it using Monte Carlo simulations. In particular,
\begin{align*}
    b_a = \frac{1}{2}\mathrm{tr}\left( \mathbf C_\Pi^{-1}\frac{\partial \mathbf C_\Pi}{\partial \theta_a} \right) &= \frac{1}{2}\mathrm{tr}\left( \mathbf C_\Pi^{-1}\frac{\partial \mathbf C_\Pi}{\partial \theta_a} \mathbf C_\Pi^{-1}\mathbf C_\Pi\right) \\
    &= \frac{1}{2}\mathrm{tr}\left( \mathbf C_\Pi^{-1}\frac{\partial \mathbf C_\Pi}{\partial \theta_a} \mathbf C_\Pi^{-1}\langle \boldsymbol d_\Pi \boldsymbol d_\Pi^\dagger \rangle \right) \\
    &= \left \langle (\mathbf C_\Pi^{-1} \boldsymbol d_\Pi)^\dagger \left(\frac{1}{2} \frac{\partial \mathbf C_\Pi}{\partial \theta_a} \right)(\mathbf C_\Pi^{-1} \boldsymbol d_\Pi)  \right\rangle = \langle E_a \rangle.
\end{align*}
To interpret the above derivation, an important observation is that it assumes that $\boldsymbol \theta_{\rm fid}$ are the true underlying parameters of the data. In that case, the mean score vanishes and $\langle \Delta \hat{\boldsymbol{\theta}}\rangle = 0$, corresponding to $\langle E_a\rangle = b_a$. We can therefore estimate $b_a$ by drawing ${\boldsymbol{n}_m}\sim{\mathcal N(0, \mathbf N_\Pi)}$ and $\boldsymbol{\delta}_m \sim \mathcal N(0, \mathbf P)$ to generate mocks
\begin{equation}
{\boldsymbol d_m = \boldsymbol{f}(\boldsymbol{\delta}_m, \boldsymbol{\theta}_{\rm fid}) + \boldsymbol{n}_m},
\end{equation}
with \textit{known true parameters} $\boldsymbol{\theta} = \boldsymbol{\theta}_{\rm fid}$. We then estimate
\begin{equation}
    b_a = \langle \,E_a({\boldsymbol{d}}_m, \boldsymbol{\theta}_{\rm fid}) \,\rangle_{\rm sims}.
\end{equation}
To generate $\boldsymbol{n}_m$, we avoid sampling from the dense matrix $\mathbf N_\Pi$ by instead drawing from $\mathcal N(0, \mathbf N)$ and applying $\mathbf \Pi$ to the result.

We use a similar approach to compute the negative Hessian $F_{ab}$. Its inverse determines the parameter covariance, so we interpret $\mathbf F$ as a Fisher matrix. Because the analysis is performed at the field level without data compression, the Fisher matrix is given by the field level Cramér–Rao bound\footnote{We provide an explicit proof in Appendix~\ref{app:ProofOptimal}.}
\begin{equation}
     F_{ab} = \frac{1}{2}\mathrm{tr}\left(\mathbf C_\Pi^{-1}\frac{\partial \mathbf{C}_\Pi}{\partial \theta_a}\mathbf C_\Pi^{-1}\frac{\partial \mathbf C_\Pi}{\partial \theta_b} \right). \label{eq:CramerRaoBound}
\end{equation}
Direct evaluation is again infeasible, so we use a stochastic estimate. As for the bias term, we rewrite
\begin{align*}
    F_{ab} = \frac{1}{2}\mathrm{tr}\left(\mathbf C_\Pi^{-1}\frac{\partial \mathbf{C}_\Pi}{\partial \theta_a}\mathbf C_\Pi^{-1}\frac{\partial \mathbf C_\Pi}{\partial \theta_b} \right) &= \mathrm{tr}\left(\mathbf C_\Pi^{-1}\frac{1}{2}\frac{\partial \mathbf{C}_\Pi}{\partial \theta_a}\mathbf C_\Pi^{-1} \frac{\partial }{\partial \theta_b^{(\rm data)}}\langle \boldsymbol{d}_\Pi \boldsymbol{d}_\Pi^\dagger \rangle  \right).
\end{align*}
Here, one derivative with respect to a parameter assumed during inference has been replaced by a derivative with respect to the true parameter underlying the data. This substitution is exact when $\boldsymbol{\theta}_{\rm fid} = \boldsymbol{\theta}^{\rm true}$. Its advantage is that ${\partial }/{\partial \theta_b^{(\rm data)}}$ can be moved outside the expectation value, because all other terms depend only on the parameters assumed during inference. Thus,
\begin{align*}
    F_{ab} &= \frac{\partial }{\partial \theta_b^{(\rm data)}} \left\langle \boldsymbol{d}_\Pi^\dagger \mathbf C_\Pi^{-1}\left(\frac{1}{2}\frac{\partial \mathbf{C}_\Pi}{\partial \theta_a} \right) \mathbf C_\Pi^{-1} \boldsymbol{d}_\Pi\right\rangle  \\
    &= \frac{\partial }{\partial \theta_b^{(\rm data)}} \left\langle \boldsymbol{w}^\dagger \left(\frac{1}{2}\frac{\partial \mathbf{C}_\Pi}{\partial \theta_a} \right) \boldsymbol{w}\right\rangle = \left\langle \frac{\partial E_a}{\partial \theta_b^{(\rm data)}} \right\rangle.
\end{align*}
We can estimate this quantity from mocks and interpret it as the {response matrix} of $E_a$. Because the minimum variance estimator ${\boldsymbol w}_m = \mathbf C_\Pi^{-1}\boldsymbol d_m$ is linear in the mock data, its change under a perturbation to the underlying parameter $\theta_b$ is
\begin{equation}
    \delta\boldsymbol{w}_{m, b} = \mathbf C_\Pi^{-1}(\partial_{\theta_b}\boldsymbol f)\boldsymbol{\delta}_m.
\end{equation}
Since $E_a$ is quadratic in $\boldsymbol{w}_m$, it follows that
\begin{equation}
    F_{ab} = \left\langle \boldsymbol{w}_{m}^\dagger\left(\frac{\partial \mathbf{C}_\Pi}{\partial \theta_a} \right)\delta\boldsymbol{w}_{m,b} \right\rangle_{\rm sims}. \label{eq:FisherApprox}
\end{equation}
The derivative is evaluated directly, avoiding finite difference errors. In practice, we use 1000 mocks to estimate $b_a$ and compute all derivatives contributing to $F_{ab}$ for each mock. The resulting stochastic noise in $\mathbf F$ and $\boldsymbol{j}$ is at the sub-percent level and can be neglected in what follows.

Using $j_a = E_a - b_a$ and the Fisher matrix in Eq.~\eqref{eq:FisherApprox}, we can now approximate the log-likelihood around the fiducial as
\begin{equation}
    \ell(\boldsymbol{\theta}) \simeq -\frac{1}{2}(\boldsymbol{\theta} - \hat{\boldsymbol{\theta}})^\dagger \,\mathbf{F}\,(\boldsymbol{\theta} - \hat{\boldsymbol{\theta}}) + \mathrm{const}, \label{eq:MUSElikelihood}
\end{equation}
where $\hat{\boldsymbol{\theta}} = \boldsymbol{\theta}_{\rm fid} + \Delta\hat{\boldsymbol{\theta}}$ and $\mathbf F \Delta\hat{\boldsymbol{\theta}} = \boldsymbol{j}$. The analysis above assumes flat priors. To include a prior $\pi(\boldsymbol{\theta})$, we sample the posterior
\begin{equation}
    \log p(\boldsymbol \theta\,|\,\boldsymbol d_\Pi) =\log \pi(\boldsymbol{\theta}) + \ell(\boldsymbol{\theta}) + \mathrm{const}.
\end{equation}
Within the Laplace approximation, we infer $(b_{1}\fnl, \fnl)$ rather than $\fnl$ alone and account for the nonlinearity induced by the coupling between $\fnl$ and $b_1$ when sampling $\log p(\boldsymbol \theta\,|\,\boldsymbol d_\Pi)$.

\subsection{Likelihood of the Summary Statistic}\label{sec:SumStat}

The MUSE analysis in the previous section depends on the data only through the estimators $E_a$. We can therefore interpret it as a form of data compression:
\begin{equation}
    p(\boldsymbol{d}\,|\,\boldsymbol{\theta}) \simeq p(\boldsymbol{E}(\boldsymbol{d}, \boldsymbol{\theta}_{\rm fid})\,|\,\boldsymbol{\theta}). \label{eq:MUSEApprox}
\end{equation}
As shown in the previous section, this compression is locally optimal.\footnote{In practice, one might imagine iteratively performing the Newton step of Eq.~\eqref{eq:NewtonStep} until some convergence criterion is reached. Asymptotically this guarantees optimality so long as the posterior is not multimodal. } Indeed, score based compression is locally optimal under suitable conditions, even for non-Gaussian likelihoods~\cite{Alsing_2018}. However, in the previous section we assumed that the resulting posterior in the parameters $\boldsymbol{\theta}$ was Gaussian. We now go beyond this approximation, which will require accurately characterizing the distribution $p(\boldsymbol{E}(\boldsymbol{d}, \boldsymbol{\theta}_{\rm fid})\,|\,\boldsymbol{\theta})$. To do so, we will first use the linear forward model to generate samples at arbitrary parameters from a single set of underlying mock fields. We then transform each estimator into a Gaussian variable and use a Gaussian copula~\cite{schmidt2006coping} to approximate their joint dependence.

In general, the distribution of each estimator is complicated. Recall that $E_a$ can be written as
\begin{equation}
    E_a(\boldsymbol{d}, \boldsymbol{\theta}_{\rm fid}) = \frac{1}{2}\boldsymbol{d}_\Pi^\dagger \mathbf Q_a\boldsymbol d_\Pi, \qquad \mathbf Q_a = \mathbf C_\Pi^{-1}\frac{\partial \mathbf C_\Pi}{\partial \theta_a}\mathbf C_\Pi^{-1}.
\end{equation}
Letting $\lambda^{(a)}_i$ denote the eigenvalues of $\mathbf {C}_\Pi^{1/2}\mathbf Q_a\mathbf {C}_\Pi^{1/2}$, it is easy to see that the above can be written as
\begin{equation}
    E_a = \frac{1}{2}\sum_i \lambda _i^{(a)}\,x_i^2, \qquad x_i \sim \mathcal{N}(0, 1^2). \label{eq:FullForm}
\end{equation}
Thus, $E_a$ is the sum over a large number of independent $\chi^2$ distributions, with weights which depend on the parameters $\boldsymbol{\theta}$ and can take either sign (since $\mathbf Q_a$ is not positive definite). Such distributions are difficult to characterize, and their likelihoods have no closed form expression~\cite{mathai1992quadratic}. A common approximation is to treat $E_a^{1/3}$ as Gaussian~\cite{abdelaty1954approximate}. 

To accurately describe the covariance of $E_a^{1/3}$, we will need to turn to mocks. We first note that it is sufficient to generate samples at $\boldsymbol{\theta}_{\rm fid}$. Since $\boldsymbol{d}$ depends on $\boldsymbol{\theta}$ linearly, the mocks at the fiducial can be analytically related to the mocks at any other choice of bias parameters. Furthermore, $E_a$ depends on $\boldsymbol{d}$ quadratically, and we can therefore trace the dependence of $E_a$ on $\boldsymbol{\theta}$ exactly. Writing $\delta \boldsymbol{\theta} = \boldsymbol{\theta} - \boldsymbol{\theta}_{\rm fid}$ and letting $E^{(m)}_a(\boldsymbol{\theta})$ denote the $m$th sample at true parameter $\boldsymbol{\theta}$, it is easy to see that $E^{(m)}_a$ takes the form
\begin{equation}
    E^{(m)}_a(\boldsymbol{\theta}) = b_a^{(m)} + F_{ab}^{(m)}\delta \theta_b + \frac{1}{2}K_{abc}^{(m)}\delta \theta_b\delta \theta_c, \label{eq:Samples}
\end{equation}
where, as in the previous section, $b_a = E_a(\boldsymbol{\theta}_{\rm fid})$, $F_{ab} = \partial E_a/\partial \theta_b$ and, similarly, we have set $K_{abc} = \partial^2 E_a/\partial \theta_b\partial \theta_c$. We denote the $m$th mock realization of the data as $\boldsymbol{d}_m$ and reintroduce the notation from the last section: $\boldsymbol{w}_m = \mathbf{C}^{-1}_\Pi \boldsymbol{d}_m, \,\,\delta\boldsymbol{w}_{m,a} = \mathbf{C}^{-1}_\Pi(\partial _a\boldsymbol{d}_m)$. With these we find 
\begin{equation}
    b_a^{(m)} = \frac{1}{2}\boldsymbol{w}_m^\dagger(\partial_a\mathbf C_\Pi)\boldsymbol{w}_m, \quad F_{ab}^{(m)} = \boldsymbol{w}_m^\dagger(\partial_a\mathbf C_\Pi)\delta\boldsymbol{w}_{m,b}, \quad K_{abc}^{(m)} = \delta\boldsymbol{w}_{m,c}^\dagger(\partial_a\mathbf C_\Pi)\delta\boldsymbol{w}_{m,b}.
\end{equation}
A few remarks are in order. We first note that at fixed $\boldsymbol{\theta}$ we only ever estimate the distribution of $\dim \boldsymbol{\theta} \simeq \mathcal{O}(10)$ variables $E_a$. Therefore, even a small number of mock realizations can accurately characterize the parameter posterior. In practice, we use 1000 mocks. A second important remark is that the dependence of the samples $E_a^{(m)}$ on the parameters $\boldsymbol{\theta}$ is smooth and the same will be true for any quantities estimated from these mocks. Indeed, the mean
\begin{equation}
    \langle E_a(\boldsymbol{\theta})\rangle = \langle b_a^{(m)}\rangle + \langle F_{ab}^{(m)}\rangle \delta \theta _b + \frac{1}{2}\langle K_{abc}^{(m)}\rangle \delta \theta _b\delta \theta_c
\end{equation}
is simply a  quadratic polynomial in the parameters and thus clearly smooth. Lastly, we note that if the quadratic contributions to $E_a$ in Eq.~\eqref{eq:Samples} are neglected and the distribution $p(\boldsymbol{E}\,|\,\boldsymbol{\theta})$ is approximated as Gaussian with parameter independent covariance, we recover the linear MUSE method presented in the previous section exactly. 

We now turn to the characterization of $p(\boldsymbol{E}\,|\,\boldsymbol{\theta})$. Following~\cite{abdelaty1954approximate}, we define the variables
\begin{equation}
    z_a(\boldsymbol{\theta}) = \sqrt[3]{\frac{E_a}{\langle E_a(\boldsymbol{\theta})\rangle}}, \label{eq:Gaussianize}
\end{equation}
which can be treated as approximately Gaussian. The Jacobian of the above transformation is easily found to be
\begin{equation}
    J_a(\boldsymbol{\theta}) = \frac{1}{3\langle E_a(\boldsymbol{\theta})\rangle}\left(\frac{E_a}{\langle E_a(\boldsymbol{\theta})\rangle} \right)^{-2/3}.
\end{equation}
Since $\{z_a\}$ are assumed to be Gaussian, their distribution is characterized entirely by the mean and covariance. We can make use of the transformation~\eqref{eq:Gaussianize} to generate samples $z_a^{(m)}(\boldsymbol{\theta})$ from the samples $E_a^{(m)}(\boldsymbol{\theta})$. We then compute the mean 
\begin{equation}
    {\boldsymbol{\mu}}(\boldsymbol{\theta}) \equiv \langle \boldsymbol{z}^{(m)}(\boldsymbol{\theta}) \rangle
\end{equation}
and covariance
\begin{equation}
    \Sigma_{ab}(\boldsymbol{\theta}) \equiv \mathrm{Cov}_m(z^{(m)}_a(\boldsymbol{\theta}), z^{(m)}_b(\boldsymbol{\theta})) \label{eq:GaussianCov}
\end{equation}
directly from the samples. The compressed likelihood $\ell(\boldsymbol{\theta}) \equiv \log p(\boldsymbol{E}\,|\,\boldsymbol{\theta})$ can now be written simply as
\begin{equation}
    \ell(\boldsymbol{\theta}) = - \frac{1}{2}(\boldsymbol{z}(\boldsymbol{\theta}) -  {\boldsymbol{\mu}}(\boldsymbol{\theta}))^\dagger\mathbf \Sigma(\boldsymbol{\theta})^{-1}(\boldsymbol{z}(\boldsymbol{\theta})- {\boldsymbol{\mu}}(\boldsymbol{\theta}))  - \frac{1}{2}\log \det \mathbf \Sigma(\boldsymbol{\theta}) + \sum_a \log |J_a(\boldsymbol{\theta})|.
    \label{eq:LogLikeSum}
\end{equation}
The covariance $\mathbf \Sigma(\boldsymbol{\theta})$ is based on mocks, however finite sample induced noise is negligible when estimating a $\mathcal O(10)$ dimensional covariance from 1000 samples. In the above, only the Jacobian $J_a(\boldsymbol{\theta}) \propto E_a(\boldsymbol{d}, \boldsymbol{\theta}_{\rm fid})^{-2/3}$ and
\begin{equation}
    z_a(\boldsymbol{d}, \boldsymbol{\theta}) = \sqrt[3]{\frac{E_a(\boldsymbol{d}, \boldsymbol{\theta}_{\rm fid})}{\langle E_a(\boldsymbol{\theta})\rangle}}
\end{equation}
depend on the data. In contrast, $\mu_a, \Sigma_{ab}$ and $ \langle E_a \rangle$ are moments taken directly from the mocks and thus carry no data dependence. We illustrate the accuracy of~\eqref{eq:LogLikeSum} in Appendix~\ref{App:Examples}. 

\section{Observational Applications}
\subsection{Optimal Weighting in \texorpdfstring{$\fnl$}{fNL} Analyses}
Traditional analyzes of galaxy surveys measure the two-point correlation function of the FKP field and relate this measurement to the theoretically predicted galaxy power spectrum. Observational effects, such as the survey window and angular and radial integral constraints, are modeled through the window matrix, while shot noise is subtracted explicitly. The covariance of the power spectrum measurements is either modeled theoretically or estimated from mocks.

\begin{figure}
    \centering
    \includegraphics[width=1\linewidth]{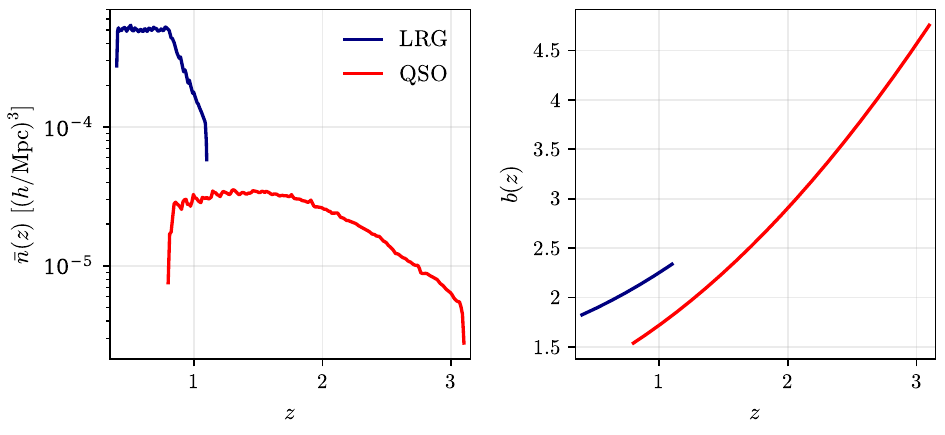}
    \caption{Assumed number density and bias evolution of the tracers considered in this forecast. The shape of $\bar n(z)$ is chosen to closely match the DR1 analysis \cite{chaussidon2025constrainingprimordialnongaussianitydesi}, but is rescaled appropriately for DR2. We assume the bias evolution is unchanged from DR1 and follows the values given in \cite{chaussidon2025constrainingprimordialnongaussianitydesi}.}
    \label{fig:bias_and_n}
\end{figure}

In the absence of a window, this prescription provides a lossless summary of the data. For a Gaussian field, all statistical information is contained in its second moment. The survey window complicates the analysis by coupling modes, an effect that is most important on scales comparable to the size of the survey. Information loss from suboptimal weighting in the power spectrum estimator is therefore most consequential for signals that dominate on large scales. A prominent example is local primordial non-Gaussianity (PNG), usually parameterized by $\fnl$. Its presence modifies the galaxy power spectrum to \cite{Dalal_2008}
\begin{equation}
    P_{gg}(k) = \left(b_1 + \fnl\frac{b_\phi}{\mathcal M(k)} \right)^2P_{mm}(k),
\end{equation}
where $\mathcal{M}(k) \propto k^2$ on large scales, so the scale dependent contribution grows as $k \rightarrow 0$. We forecast the improvement obtainable through optimal weighting in a DESI DR2-like survey of luminous red galaxies (LRGs) and quasars (QSOs), using their assumed footprints, bias evolution, redshift ranges, and number densities. Figure~\ref{fig:bias_and_n} shows the assumed number density and bias evolution, which are chosen to closely resemble the DESI DR2 specifications. 

\begin{figure}
    \centering
    \includegraphics[width=0.9\linewidth]{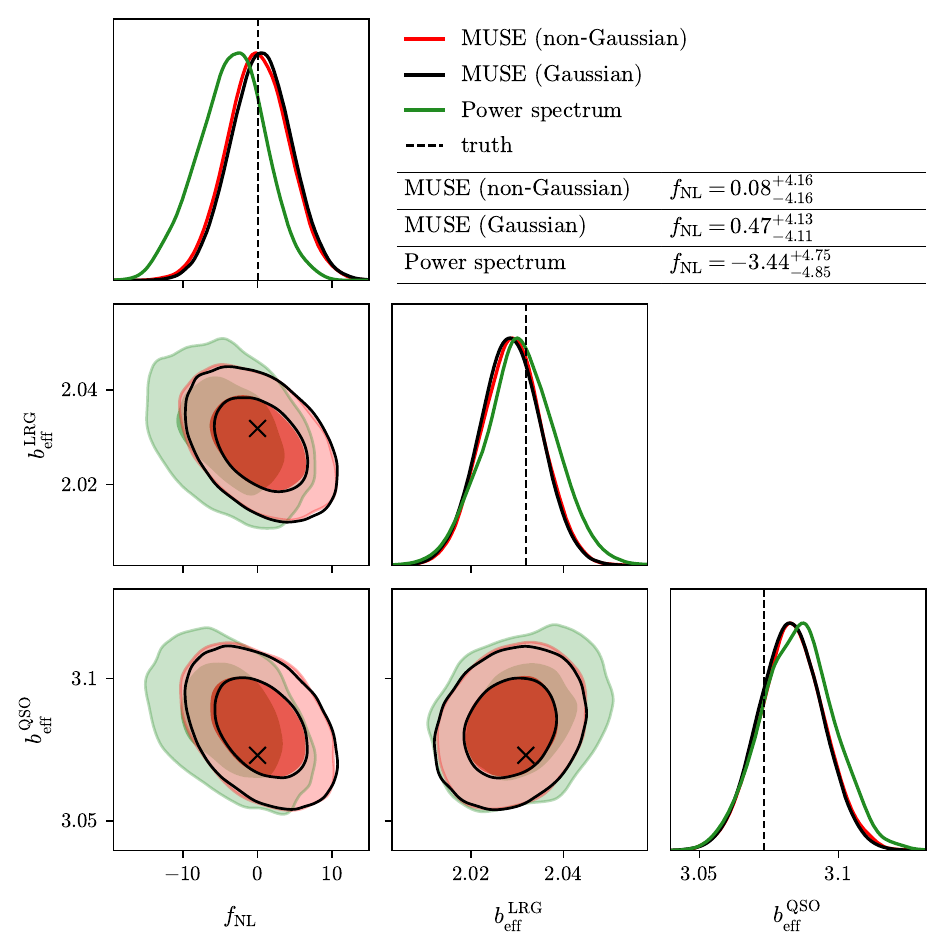}
    \caption{Posterior for the bias parameters obtained using the field level likelihood (MUSE) and the power spectrum analysis. MUSE (Gaussian) denotes the Gaussian approximation of Sec.~\ref{sec:Laplace} and MUSE (non-Gaussian) the parameter dependent likelihood of the summary statistic derived in Sec.~\ref{sec:SumStat}. Through optimal weighting of the data, the field level analysis tightens the constraints on $\fnl$. The non-Gaussian corrections to the MUSE likelihood introduced in Sec.~\ref{sec:SumStat} have only a percent level effect. }
    \label{fig:MTcorner}
\end{figure}

Following \cite{Abdul_Karim_2025}, we assume angular footprints of approximately $A_{\rm LRG} \sim 10\,000\,\mathrm{deg}^2$ and $A_{\rm QSO} \sim 11\,000\,\mathrm{deg}^2$, divided between the Northern and Southern Galactic Caps (NGC and SGC).\footnote{We approximate the angular mask by restricting the range of the polar angle.} To reduce computational cost, we analyze only the NGC and rescale the resulting likelihood to forecast constraints from a joint analysis of both caps. This approximation neglects correlations between the caps, as is also done in the DESI analysis~\cite{chaussidon2025constrainingprimordialnongaussianitydesi}. We model both tracer samples in a cube of side length $L = 7.5\,\mathrm{Gpc}/h$ on a mesh with $N = 256$ cells per side. This accommodates both samples and gives $k_{\rm Ny} = 0.11\,h/\mathrm{Mpc} > k_{\max}$. Following the DESI DR1 analysis \cite{chaussidon2025constrainingprimordialnongaussianitydesi}, we restrict the analysis to $k < k_{\rm max} = 0.08\,h/\mathrm{Mpc}$ and also impose a cut on large scales. Somewhat more optimistically than DESI, we use $k_{\min} = 10^{-3}\,h/\mathrm{Mpc}$ for the QSO sample and $k_{\min} = 3\cdot 10^{-3}\,h/\mathrm{Mpc}$ for the LRGs. We will discuss the presence of angular systematics, which may complicate this choice in practice, in the subsequent section. 

\begin{figure}
    \centering
    \includegraphics[width=1\linewidth]{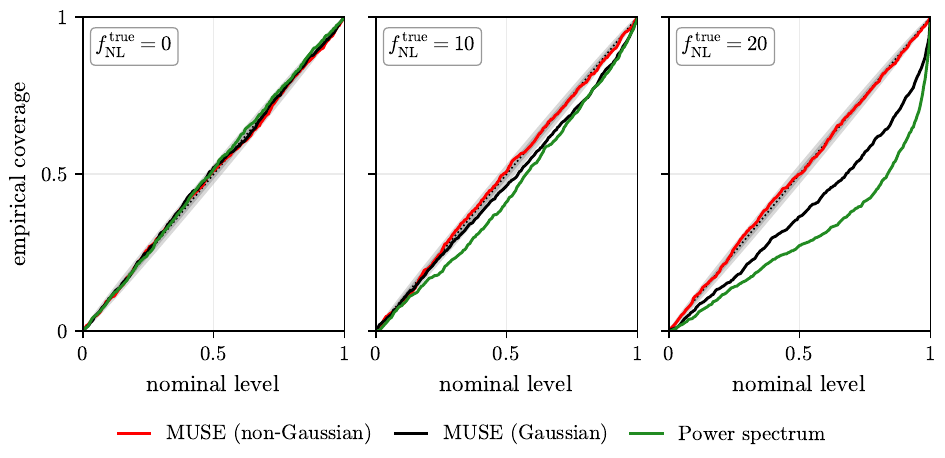}
    \caption{Coverage plot for the $\fnl$ posterior obtained using various methods. Well calibrated posteriors lie on the diagonal, and we indicate the $1\sigma$ and $2\sigma$ binomial scatter in gray. At the fiducial value used to estimate the corresponding covariances, all methods are well calibrated and deliver reliable errors. However, when $\fnl^{\rm true} \neq 0$, methods that rely on a parameter independent covariance become unreliable. The fully parameter dependent prescription presented in Sec.~\ref{sec:SumStat} remains unbiased across a wide range of true $\fnl$ values.}
    \label{fig:coverage}
\end{figure}

To compare the power spectrum and field level analyses, we use the same bias parameters $(b_1, \fnl, \Sigma_s)$ for both. The field level model is described in Sec.~\ref{sec:Theory1}. For the power spectrum analysis, we follow the DESI prescription and write
\begin{equation}
    P_{gg}(k, \mu) = \left(b_1 + f(z_{\rm eff})\mu^2+ \fnl \frac{2\delta_c(b_1 - p)}{\mathcal M(k, z_{\rm eff})}\right)^2\frac{P_{mm}(k, z_{\rm eff})}{\left(1 + \tfrac{1}{2}(k\mu\Sigma_s)^2 \right)^2},
\end{equation}
with $p = 1$ for LRGs and $p = 1.4$ for QSOs, as suggested by \cite{fondi2026assemblybiaslocalprimordial}. We use the optimal quadratic estimator (OQE) prescription for the effective redshift and per galaxy weights \cite{chaussidon2025constrainingprimordialnongaussianitydesi, Castorina_2019_OQE}, which improves on standard FKP weighting by approximately $10$--$15\%$. We measure all power spectra with \texttt{pypower},\footnote{\url{https://github.com/cosmodesi/pypower}} using a custom window matrix implementation to account for the radial and angular integral constraints described in Appendix~\ref{app:IC}. We estimate the covariance of the power spectrum estimator from $1000$ Gaussian mocks generated with the forward model from Sec.~\ref{sec:Theory1}. Thus, the field level and power spectrum forecasts use the same underlying statistical model.

To analyze the LRG and QSO samples jointly, including their cross-correlation, we extend the setup of Sec.~\ref{sec:Theory1}. For data vectors $\boldsymbol{d}_{\rm LRG}$ and $\boldsymbol{d}_{\rm QSO}$, and corresponding forward models $\boldsymbol{f}_{\rm LRG}$ and $\boldsymbol{f}_{\rm QSO}$, we write
\begin{equation}
    \boldsymbol{d} \equiv \begin{pmatrix}
        \boldsymbol{d}_{\rm LRG} \\ \boldsymbol{d}_{\rm QSO}
    \end{pmatrix} = \begin{pmatrix}
        \boldsymbol{f}_{\rm LRG}(\boldsymbol{\delta}) \\ \boldsymbol{f}_{\rm QSO}(\boldsymbol{\delta}) 
    \end{pmatrix} + \begin{pmatrix}
        \boldsymbol{n}_{\rm LRG} \\ \boldsymbol{n}_{\rm QSO}
    \end{pmatrix}.
\end{equation}
Describing the data through a shared matter field $\boldsymbol{\delta}$ automatically accounts for the cross-covariance between the two tracers. Assuming no cross-shot noise \cite{chaussidon2025constrainingprimordialnongaussianitydesi}, the noise covariance is block diagonal:
\begin{equation}
    \mathbf N = \begin{pmatrix}
        \mathbf N_{\rm LRG} & \mathbf 0 \\
        \mathbf 0 & \mathbf N_{\rm QSO}
    \end{pmatrix}.
\end{equation}
The remaining analysis follows the procedure described in the preceding sections, with the parameter vector now reading $\boldsymbol{\theta} = (b_1^{\rm LRG}, \Sigma_s^{\rm LRG}, b_1^{\rm QSO}, \Sigma_s^{\rm QSO}, \fnl)$. 

We compare the posterior from the field level analysis, using either the Gaussian approximation or the statistical description of Sec.~\ref{sec:SumStat}, with the constraints obtained from the power spectrum analysis across $1000$ mocks. The posterior for a representative sample is shown in Fig.~\ref{fig:MTcorner}. Through optimal weighting of the data, the field level analysis achieves $14\%$ tighter constraints than the traditional power spectrum analysis. The posterior of the bias parameters is already close to Gaussian, so the methods presented in Sec.~\ref{sec:Laplace} and Sec.~\ref{sec:SumStat} differ only marginally.

To assess the calibration of the $\fnl$ constraints from the field level and power spectrum analyses, we perform a coverage test. For a given true value $f_{\rm NL}^{\rm true}$, we draw $N = 1000$ test mocks that are independent of the mocks used to calibrate the field level likelihood and estimate the power spectrum covariance. For each test mock $m$, method $M$ and nominal level $\alpha$, we compute the central credible interval of the marginal posterior of $\fnl$,
\begin{equation}
    {\rm CI}^{M}_{\alpha}(\boldsymbol{d}_m) = \big[\, q^{M}_{(1-\alpha)/2}(\boldsymbol{d}_m),\; q^{M}_{(1+\alpha)/2}(\boldsymbol{d}_m) \,\big] ,
\end{equation}
where $q^{M}_{p}(\boldsymbol{d}_m)$ is the $p$-quantile of the posterior for the $m$th mock. We define the \textit{empirical coverage} as the fraction of mocks whose interval contains the true value,
\begin{equation}
    C_M(\alpha) = \frac{1}{N}\sum_{m=1}^{N} \begin{cases}
         1 & f_{\rm NL}^{\rm true} \in {\rm CI}^{M}_{\alpha} (\boldsymbol{d}_m) \\
         0 & \mathrm{else}
    \end{cases} \bigg\}.
\end{equation}
For a calibrated posterior, $\langle C_M(\alpha) \rangle = \alpha$. A curve below the diagonal, $C_M(\alpha) < \alpha$, means that the credible intervals contain the truth less often than their nominal probability. That is, the posterior is overconfident or biased. A curve above the diagonal means that the posterior is conservative. We show the coverage for three true values of $\fnl$ in Fig.~\ref{fig:coverage}. As is visible from the figure, all methods are well calibrated at the fiducial value $\fnl^{\rm true} = 0$. However, when $\fnl^{\rm true} \neq 0$, methods that use a parameter independent covariance can become overconfident.

\begin{figure}
    \centering
    \includegraphics[width=1\linewidth]{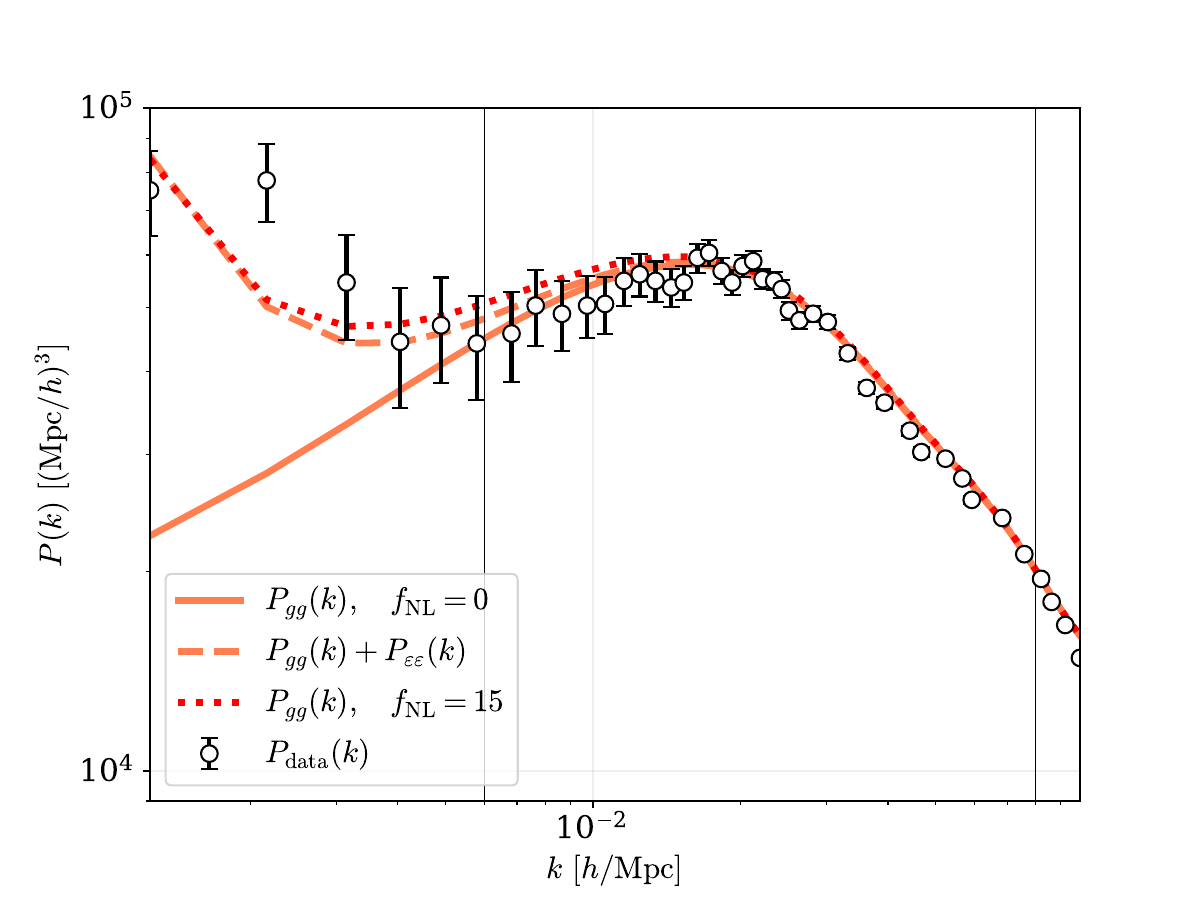}
    \caption{An angular contaminant with a power-law spectrum can closely resemble the contribution of local primordial non-Gaussianity to the power spectrum monopole. The points show the DESI DR1 LRG monopole from~\cite{rosadomarín2026localprimordialnongaussianityusing}. The vertical lines indicate the scale range used in the fiducial DESI analyses~\cite{chaussidon2025constrainingprimordialnongaussianitydesi,rosadomarín2026localprimordialnongaussianityusing}. Restricting the analysis to this range reduces sensitivity to contamination, but discards information from the largest scales.}
    \label{fig:dust_vs_fnl}
\end{figure}

\subsection{Unknown Angular Contaminants}

Known angular systematics can be mitigated by projecting out a finite set of templates, as described in Sec.~\ref{sec:IC}. Real data may also contain angular contaminants for which no reliable template is available~\cite{rosadomarín2026localprimordialnongaussianityusing}. If these contaminants are not modeled, they can bias the analysis on large scales. This is particularly important for local primordial non-Gaussianity. As illustrated in Fig.~\ref{fig:dust_vs_fnl}, an angular contaminant can produce a contribution to the power spectrum monopole that closely resembles a primordial signal. Raising $k_{\min}$ reduces sensitivity to this contamination, but also removes the large scale modes that are especially informative about $\fnl$. In addition, the analysis becomes sensitive to an arbitrarily chosen scale cut.

At the field level, we can instead include an unknown angular field $\boldsymbol{\varepsilon}$ in the data model. For the projected data, we write
\begin{equation}
    \boldsymbol{d}_{\Pi}
    = \boldsymbol{f}(\boldsymbol{\delta},\boldsymbol{\theta})
    + \boldsymbol{g}(\boldsymbol{\varepsilon})
    + \boldsymbol{n}_{\Pi},
\end{equation}
where $\boldsymbol{g}$ maps the angular field into the three dimensional data. This model treats the contaminant as additive. To first order, a multiplicative modulation of the galaxy number density can equivalently be treated as an additive contribution~\cite{Weaverdyck_2021}:
\begin{equation}
    n(\boldsymbol{x})
    = \bar n(\boldsymbol{x})
      (1+\delta_g(\boldsymbol{x}))
      (1+\varepsilon(\hat{\boldsymbol{x}}))
    \simeq
    \bar n(\boldsymbol{x})
      (1+\delta_g(\boldsymbol{x})
       +\varepsilon(\hat{\boldsymbol{x}})).
    \label{eq:ExpansionDust}
\end{equation}
An angular field has no independent radial structure. In the idealized flat sky limit, before applying the survey window, its power spectrum is confined to $k_r=0$, i.e.
\begin{equation}
\langle\varepsilon(\boldsymbol{k})\varepsilon(\boldsymbol{k})^*\rangle \propto \delta(\mu), \qquad \mu = k_r/k.
\end{equation}
The finite radial depth of the survey broadens this contribution, but does not make it equivalent to an unrestricted three dimensional field. Marginalizing over $\boldsymbol{\varepsilon}$ can therefore retain radial modes that a simple cut $k > k_{\min}$ would discard.

The methods of Sec.~\ref{sec:MUSE} extend directly to this model. Assuming that $\boldsymbol{\varepsilon}$ is Gaussian and independent of the matter field and noise, marginalizing over it gives
\begin{equation}
    \mathbf C_{\Pi}
    = \boldsymbol{f}\mathbf P\boldsymbol{f}^{\dagger}
    + \boldsymbol{g}
      \langle\boldsymbol{\varepsilon}
      \boldsymbol{\varepsilon}^{\dagger}\rangle
      \boldsymbol{g}^{\dagger}
    + \mathbf N_{\Pi}.
    \label{eq:CovarianceDust}
\end{equation}
If the angular power spectrum is known, the contaminant simply widens the data covariance. Here, we instead allow its broadband power to be inferred jointly with the cosmological parameters. This accounts for the correlations between the contaminant amplitude and cosmological signals.

\begin{figure}
    \centering
    \includegraphics[width=1\linewidth]{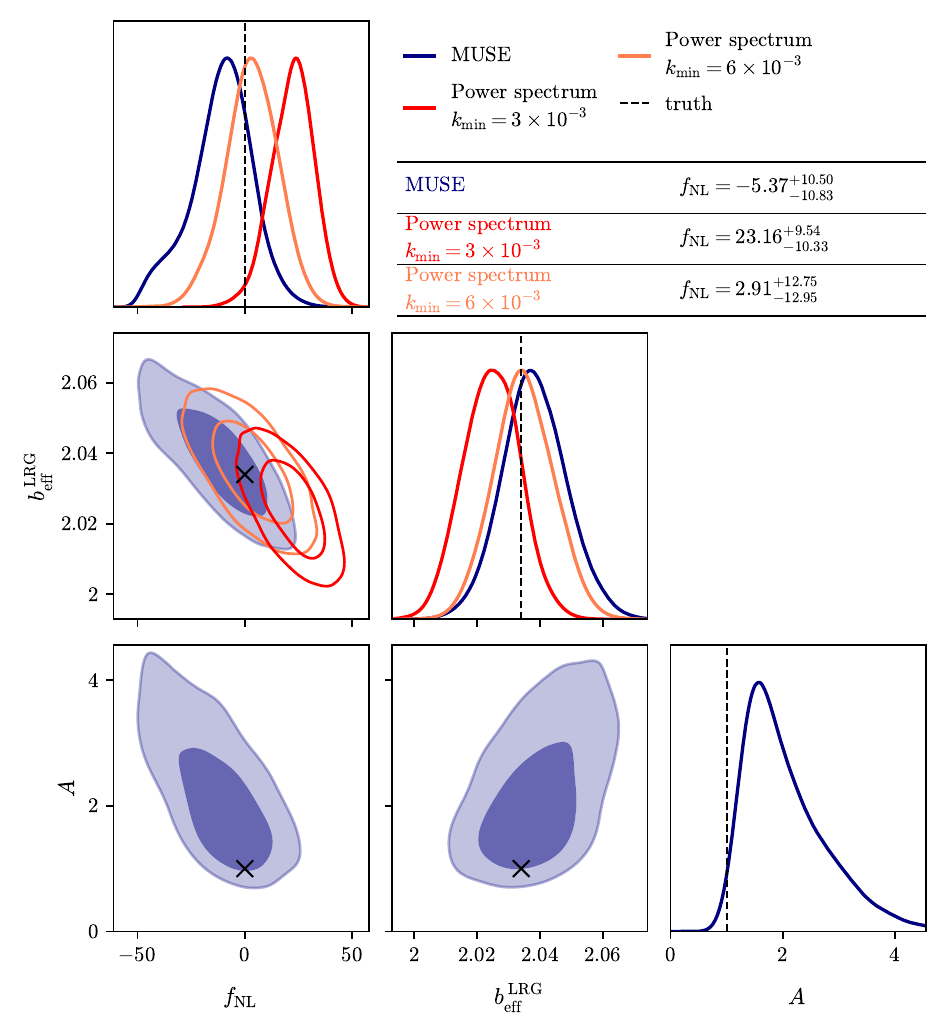}
    \caption{Posterior of the LRG bias parameters $(\fnl, b_1)$ and the amplitude of the angular contaminant $A$, which is only constrained at the field level. Field level analyses can marginalize over unknown angular contaminants, while power spectrum analyses become unbiased only after introducing appropriate scale cuts. Marginalizing over an arbitrary angular contaminant with free amplitude increases the uncertainty by 20\% relative to unmarginalized error of 9. 
    The fiducial DESI choice, ${k_{\min} = 6 \cdot 10^{-3}\,h/\mathrm{Mpc}}$, widens the errors by an additional 20\% relative to marginalized MUSE, while also introducing an arbitrary parameter into the analysis. The field level errors include additional variance from the stochastic Gaussian contaminant in Eq.~\eqref{eq:CovarianceDust}, which is absent from the power spectrum covariance, making the power spectrum errors overconfident.}
    \label{fig:dust}
\end{figure}

 Assuming a fiducial angular power spectrum $C_\ell^\varepsilon$,\footnote{The shape of the angular power spectrum must be taken from a physically informed model, such as a power law~\cite{2016_PLANCK_DUST}. Alternatively, the binned broadband parameters of this spectrum could be inferred individually. Neither choice has a significant impact on the $\fnl$ posterior.} such that
\begin{equation}
    \langle\varepsilon_{\ell m}
    \varepsilon_{\ell' m'}^{*}\rangle
    = C_\ell^\varepsilon\,\delta_{\ell\ell'}\delta_{mm'},
\end{equation}
we can parameterize the overall amplitude of the contaminant signal, $A$, as part of the forward model $\boldsymbol{g}$. This has the advantage that the data remain linear in the inferred parameters. The response is then
\begin{equation}
    [\boldsymbol{g}(\boldsymbol{\varepsilon})](\boldsymbol{x})
    =
    \mathbf \Pi\,\left(
    \bar n(\boldsymbol{x})
    A\,\sum_{\ell m}
    Y_{\ell m}(\hat{\boldsymbol{x}})
    \,\varepsilon_{\ell m} \right).
\end{equation}
The factor $\bar n(\boldsymbol{x})$ follows from Eq.~\eqref{eq:ExpansionDust}, while $\mathbf \Pi$ applies the same mode restrictions and integral constraints as in the cosmological forward model. The spherical harmonics can be evaluated on a \texttt{HEALPix} grid~\cite{gorski1999healpixprimer} with $N_{\rm side}=64$. The operator $\boldsymbol{g}$ maps a two dimensional set of harmonic coefficients into the three dimensional data. It need not be represented as a dense matrix and can instead be evaluated efficiently in the above form with $\ell_{\rm max} = 64$.

The optimal quadratic estimator for $A$ follows directly from the formalism presented in Sec.~\ref{sec:MUSE}. Defining the Wiener reconstruction
\begin{equation}
    \hat{\boldsymbol{\varepsilon}}
    = \mathbf C^\varepsilon\boldsymbol{g}^{\dagger}\mathbf C_\Pi^{-1}\boldsymbol{d}_\Pi,
\end{equation}
where $[\mathbf C^\varepsilon \boldsymbol{\varepsilon}]_{\ell m} = C^\varepsilon_{\ell}\,\varepsilon_{\ell m}$, we obtain
\begin{equation}
    E_{A} = \sum_{\ell m}\frac{|\hat\varepsilon_{\ell m}|^2}{C^\varepsilon_\ell}.
\end{equation}
The amplitude $A$ can also be interpreted as the response of the survey to angular modes with power spectrum $C_\ell^{\varepsilon}$. Within this interpretation, it may be natural to allow the amplitude to vary with redshift.

We present the results of marginalizing over an unknown angular systematic in Fig.~\ref{fig:dust}. The spectrum of the injected contaminant was chosen to closely match the excess power seen in the monopole of the DESI DR1 LRG sample~\cite{rosadomarín2026localprimordialnongaussianityusing}. At the fiducial choice $k_{\min} = 6 \cdot 10^{-3}\,h/\mathrm{Mpc}$, the analysis based on the power spectrum 
becomes unbiased. However, this requires specifying an arbitrary scale cut and discards uncontaminated information from scales $k < k_{\min}$. The field level analysis can use all scales present in the data while remaining robust to angular contamination and accurately capturing the degeneracy between the contaminant amplitude and the primordial signal. It achieves 20\% lower error on $f_{\rm NL}$ than the fiducial power
spectrum analysis. 

\section{Conclusions}
\label{sec:Conclusion}

We have presented a field level analysis of galaxy surveys on large scales. Our approach directly models the observed density field, including the survey window, integral constraints and shot noise, before compressing the data into a small set of optimal quadratic estimators. This addresses the central challenge of optimally extracting the large scale information relevant for local primordial non-Gaussianity in the presence of a survey window, which couples modes and makes conventional power spectrum estimators suboptimal.

The resulting MUSE compression is locally optimal under the Gaussian field model. In particular, the compressed estimators retain the Fisher information of the field level likelihood while reducing the inference problem to a low dimensional statistical description. For the DESI DR2 LRG and QSO samples, optimal field level weighting improves the forecasted constraint on $\fnl$ by $14\%$ relative to the power spectrum analysis. We further showed that the Gaussian approximation is accurate near the fiducial model, but can become overconfident away from it when the covariance is held fixed. Modeling the parameter dependence of the compressed likelihood restores calibrated posteriors across a wide range of $\fnl$ values.

The field level formulation also treats observational effects more naturally than a power spectrum analysis. The survey window and radial and angular integral constraints are incorporated directly into the forward model. Unknown angular contaminants can be marginalized over at the field level without imposing an arbitrary large scale cut, while retaining the radial modes that remain cosmologically informative. This can lead 
to additional gains relative to the power 
spectrum analysis. 
Additionally, the analysis remains computationally manageable. Generating $100$ mocks and evaluating the quantities required for the field level likelihood takes approximately $4$ GPU hours on one NERSC GPU node with four A100 GPUs. This is substantially less expensive than directly sampling the high dimensional field level posterior. 

 The present analysis is restricted to a linear bias model and to scales where the Gaussian description of the density field is accurate. A natural extension is to incorporate nonlinear bias and forward models within the same framework~\cite{Seljak_2017}, which could recover information beyond the power spectrum. For example, in DESI DR1 data, adding the QSO bispectrum monopole to the QSO power spectrum improves the local PNG constraint by about $15\%$~\cite{Chudaykin_2025}. 

We plan to apply the presented method to the DESI DR2 data, where we expect to improve the LRG and QSO error by 15-20\%. This will require validating the forward model, the treatment of observational systematics and the parameter dependent likelihood on realistic mocks. A natural byproduct of the field level analysis is the Wiener reconstruction of the matter field. Recent work shows that splitting tracers by their reconstructed large scale dark matter environment can provide a factor of $2$--$3$ improvement in the LRG and 20\% in QSO constraint on $\fnl$~\cite{adolff2026itsenvironmentlocalfrm}. Applying this method to data will also require carefully modeling the covariance between the low  and high density samples, including their non-Poisson stochasticity and the variance introduced by using the observed galaxy field to define the split. 

\acknowledgments

We would like to thank Edmond Chaussidon, Richard Feder and Noah Weaverdyck for many insightful discussions. This research used resources of the National Energy Research Scientific Computing Center (NERSC), a Department of Energy Office of Science User Facility. We acknowledge the use of the Anthropic large language models \texttt{Claude Fable 5.1} and \texttt{Claude Opus 5.5} during code implementation. This work is supported by NSF CDSE grant number AST-2408026 and NASA TCAN grant number 80NSSC24K0101.

\bibliographystyle{JHEP}
\bibliography{bibliography}

@article{Feldman_1994,
   title="{Power-spectrum analysis of three-dimensional redshift surveys}",
   volume={426},
   ISSN={1538-4357},
   url={http://dx.doi.org/10.1086/174036},
   DOI={10.1086/174036},
   journal={The Astrophysical Journal},
   publisher={American Astronomical Society},
   author={Feldman, Hume A. and Kaiser, Nick and Peacock, John A.},
   year={1994},
   month=May, pages={23} }

@article{ross2024constructionlargescalestructurecatalogs,
      title="{The Construction of Large-scale Structure Catalogs for the Dark Energy Spectroscopic Instrument}", 
      author={A. J. Ross and J. Aguilar and S. Ahlen and S. Alam and A. Anand and S. Bailey and D. Bianchi and S. Brieden and D. Brooks and E. Burtin and A. Carnero Rosell and E. Chaussidon and T. Claybaugh and S. Cole and K. Dawson and A. de la Macorra and A. de Mattia and Arjun Dey and Biprateep Dey and P. Doel and K. Fanning and S. Ferraro and J. Ereza and A. Font-Ribera and J. E. Forero-Romero and E. Gaztañaga and H. Gil-Marín and S. Gontcho A Gontcho and A. X. Gonzalez-Morales and J. Guy and C. Hahn and S. Heydenreich and K. Honscheid and C. Howlett and M. Ishak and T. Karim and D. Kirkby and T. Kisner and H. Kong and A. Kremin and A. Krolewski and A. Lambert and M. Landriau and J. Lasker and L. Le Guillou and M. E. Levi and M. Manera and P. Martini and P. McDonald and A. Meisner and R. Miquel and J. Moon and J. Moustakas and A. Muñoz-Gutiérrez and A. D. Myers and S. Nadathur and L. Napolitano and J. A. Newman and J. Nie and G. Niz and N. Palanque-Delabrouille and W. J. Percival and C. Poppett and F. Prada and A. Raichoor and C. Ravoux and M. Rezaie and A. Rosado-Marin and G. Rossi and L. Samushia and E. Sanchez and E. F. Schlafly and D. Schlegel and H. Seo and A. Smith and D. Sprayberry and G. Tarlé and D. Valcin and M. Vargas-Magaña and B. A. Weaver and M. Wilson and J. Yu and P. Zarrouk and C. Zhao and R. Zhou and H. Zou},
      year={2025},
      eprint={2405.16593},
      archivePrefix={arXiv},
      primaryClass={astro-ph.CO},
      journal={Journal of Cosmology and Astroparticle Physics},
      volume={01},
      pages={125},
      doi={10.1088/1475-7516/2025/01/125},
      url={https://arxiv.org/abs/2405.16593}, 
}

@article{Ross_2012,
   title="{The clustering of galaxies in the SDSS-III Baryon Oscillation Spectroscopic Survey: analysis of potential systematics: Systematic analysis of BOSS 3D clustering}",
   volume={424},
   ISSN={0035-8711},
   url={http://dx.doi.org/10.1111/j.1365-2966.2012.21235.x},
   DOI={10.1111/j.1365-2966.2012.21235.x},
   number={1},
   journal={Monthly Notices of the Royal Astronomical Society},
   publisher={Oxford University Press (OUP)},
   author={Ross, Ashley J. and Percival, Will J. and Sánchez, Ariel G. and Samushia, Lado and Ho, Shirley and Kazin, Eyal and Manera, Marc and Reid, Beth and White, Martin and Tojeiro, Rita and McBride, Cameron K. and Xu, Xiaoying and Wake, David A. and Strauss, Michael A. and Montesano, Francesco and Swanson, Molly E. C. and Bailey, Stephen and Bolton, Adam S. and Dorta, Antonio Montero and Eisenstein, Daniel J. and Guo, Hong and Hamilton, Jean-Christophe and Nichol, Robert C. and Padmanabhan, Nikhil and Prada, Francisco and Schlegel, David J. and Magaña, Mariana Vargas and Zehavi, Idit and Blanton, Michael and Bizyaev, Dmitry and Brewington, Howard and Cuesta, Antonio J. and Malanushenko, Elena and Malanushenko, Viktor and Oravetz, Daniel and Parejko, John and Pan, Kaike and Schneider, Donald P. and Shelden, Alaina and Simmons, Audrey and Snedden, Stephanie and Zhao, Gong-bo},
   year={2012},
   month=jun, pages={564--590} }

@article{Bernardeau_2002,
   title="{Large-scale structure of the Universe and cosmological perturbation theory}",
   volume={367},
   ISSN={0370-1573},
   url={http://dx.doi.org/10.1016/S0370-1573(02)00135-7},
   DOI={10.1016/s0370-1573(02)00135-7},
   number={1-3},
   journal={Physics Reports},
   publisher={Elsevier BV},
   author={Bernardeau, F. and Colombi, S. and Gaztañaga, E. and Scoccimarro, R.},
   year={2002},
   month=sep, pages={1--248} }

@article{Beutler_2019,
   title="{Interpreting measurements of the anisotropic galaxy power spectrum}",
   volume={03},
   ISSN={1475-7516},
   url={http://dx.doi.org/10.1088/1475-7516/2019/03/040},
   DOI={10.1088/1475-7516/2019/03/040},
   number={03},
   journal={Journal of Cosmology and Astroparticle Physics},
   publisher={IOP Publishing},
   author={Beutler, Florian and Castorina, Emanuele and Zhang, Pierre},
   year={2019},
   month=Mar, pages={040} }

@article{Horowitz_2019,
   title="{Efficient optimal reconstruction of linear fields and band-powers from cosmological data}",
   volume={10},
   ISSN={1475-7516},
   url={http://dx.doi.org/10.1088/1475-7516/2019/10/035},
   DOI={10.1088/1475-7516/2019/10/035},
   number={10},
   journal={Journal of Cosmology and Astroparticle Physics},
   publisher={IOP Publishing},
   author={Horowitz, B. and Seljak, U. and Aslanyan, G.},
   year={2019},
   month=Oct, pages={035} }

@article{Seljak_2017,
   title="{Towards optimal extraction of cosmological information from nonlinear data}",
   volume={12},
   ISSN={1475-7516},
   url={http://dx.doi.org/10.1088/1475-7516/2017/12/009},
   DOI={10.1088/1475-7516/2017/12/009},
   number={12},
   journal={Journal of Cosmology and Astroparticle Physics},
   publisher={IOP Publishing},
   author={Seljak, Uroš and Aslanyan, Grigor and Feng, Yu and Modi, Chirag},
   year={2017},
   month=Dec, pages={009} }

@article{Seljak_1998,
doi = {10.1086/306019},
url = {https://doi.org/10.1086/306019},
year = {1998},
month = {aug},
publisher = {},
volume = {503},
number = {2},
pages = {492},
author = {Seljak, Uroš},
title = "{Cosmography and Power Spectrum Estimation: A Unified Approach}",
journal = {The Astrophysical Journal}
}

@article{Millea_2022,
   title="{Marginal unbiased score expansion and application to CMB lensing}",
   volume={105},
   ISSN={2470-0029},
   url={http://dx.doi.org/10.1103/PhysRevD.105.103531},
   DOI={10.1103/physrevd.105.103531},
   number={10},
   journal={Physical Review D},
   publisher={American Physical Society (APS)},
   author={Millea, Marius and Seljak, Uroš},
   year={2022},
   month=May, pages={103531} }

@article{Castorina_2018,
   title="{Beyond the plane-parallel approximation for redshift surveys}",
   ISSN={1365-2966},
   url={http://dx.doi.org/10.1093/mnras/sty410},
   DOI={10.1093/mnras/sty410},
   journal={Monthly Notices of the Royal Astronomical Society},
   publisher={Oxford University Press (OUP)},
   author={Castorina, Emanuele and White, Martin},
   year={2018},
   month=Feb, volume={476}, pages={4403--4417} }

@article{Desjacques_2018,
   title="{Large-scale galaxy bias}",
   volume={733},
   ISSN={0370-1573},
   url={http://dx.doi.org/10.1016/j.physrep.2017.12.002},
   DOI={10.1016/j.physrep.2017.12.002},
   journal={Physics Reports},
   publisher={Elsevier BV},
   author={Desjacques, Vincent and Jeong, Donghui and Schmidt, Fabian},
   year={2018},
   month=Feb, pages={1--193} }

@article{Seljak_2011,
   title="{Distribution function approach to redshift space distortions}",
   volume={11},
   ISSN={1475-7516},
   url={http://dx.doi.org/10.1088/1475-7516/2011/11/039},
   DOI={10.1088/1475-7516/2011/11/039},
   number={11},
   journal={Journal of Cosmology and Astroparticle Physics},
   publisher={IOP Publishing},
   author={Seljak, Uroš and McDonald, Patrick},
   year={2011},
   month=Nov, pages={039} }

@article{Jackson_1972,
   title="{A Critique of Rees’s Theory of Primordial Gravitational Radiation}",
   volume={156},
   ISSN={1365-2966},
   url={http://dx.doi.org/10.1093/mnras/156.1.1P},
   DOI={10.1093/mnras/156.1.1p},
   number={1},
   journal={Monthly Notices of the Royal Astronomical Society},
   publisher={Oxford University Press (OUP)},
   author={Jackson, J. C.},
   year={1972},
   month=Feb, pages={1P--5P} }

@article{Laurent_2017,
   title="{Clustering of quasars in SDSS-IV eBOSS: study of potential systematics and bias determination}",
   volume={07},
   ISSN={1475-7516},
   url={http://dx.doi.org/10.1088/1475-7516/2017/07/017},
   DOI={10.1088/1475-7516/2017/07/017},
   number={07},
   journal={Journal of Cosmology and Astroparticle Physics},
   publisher={IOP Publishing},
   author={Laurent, Pierre and Eftekharzadeh, Sarah and Goff, Jean-Marc Le and Myers, Adam and Burtin, Etienne and White, Martin and Ross, Ashley J. and Tinker, Jeremy and Tojeiro, Rita and Bautista, Julian and Brinkmann, Jonathan and Comparat, Johan and Dawson, Kyle and Bourboux, Hélion du Mas des and Kneib, Jean-Paul and McGreer, Ian D. and Palanque-Delabrouille, Nathalie and Percival, Will J. and Prada, Francisco and Rossi, Graziano and Schneider, Donald P. and Weinberg, David and Yèche, Christophe and Zarrouk, Pauline and Zhao, Gong-Bo},
   year={2017},
   month=jul, pages={017} }

@article{chaussidon2025constrainingprimordialnongaussianitydesi,
      title="{Constraining primordial non-Gaussianity with DESI 2024 LRG and QSO samples}", 
      author={E. Chaussidon and C. Yèche and A. de Mattia and C. Payerne and P. McDonald and A. J. Ross and S. Ahlen and D. Bianchi and D. Brooks and E. Burtin and T. Claybaugh and A. de la Macorra and P. Doel and S. Ferraro and A. Font-Ribera and J. E. Forero-Romero and E. Gaztañaga and H. Gil-Marín and S. Gontcho A Gontcho and G. Gutierrez and J. Guy and K. Honscheid and C. Howlett and D. Huterer and R. Kehoe and D. Kirkby and T. Kisner and A. Kremin and L. Le Guillou and M. E. Levi and M. Manera and A. Meisner and R. Miquel and J. Moustakas and J. A. Newman and G. Niz and N. Palanque-Delabrouille and W. J. Percival and F. Prada and I. Pérez-Ràfols and C. Ravoux and G. Rossi and E. Sanchez and D. Schlegel and M. Schubnell and H. Seo and D. Sprayberry and G. Tarlé and M. Vargas-Magaña and B. A. Weaver and C. Zhao and H. Zou},
      year={2025},
      eprint={2411.17623},
      archivePrefix={arXiv},
      primaryClass={astro-ph.CO},
      journal={Journal of Cosmology and Astroparticle Physics},
      volume={06},
      pages={029},
      doi={10.1088/1475-7516/2025/06/029},
      url={https://arxiv.org/abs/2411.17623}, 
}

@article{de_Mattia_2019,
   title="{Integral constraints in spectroscopic surveys}",
   volume={08},
   ISSN={1475-7516},
   url={http://dx.doi.org/10.1088/1475-7516/2019/08/036},
   DOI={10.1088/1475-7516/2019/08/036},
   number={08},
   journal={Journal of Cosmology and Astroparticle Physics},
   publisher={IOP Publishing},
   author={de Mattia, Arnaud and Ruhlmann-Kleider, Vanina},
   year={2019},
   month=Aug, pages={036} }

@ARTICLE{1991MNRAS,
       author = {{Peacock}, J.~A. and {Nicholson}, D.},
        title = "{The large-scale clustering of radio galaxies}",
      journal = {Monthly Notices of the Royal Astronomical Society},
         year = 1991,
        month = nov,
       volume = {253},
        pages = {307-319},
          doi = {10.1093/mnras/253.2.307},
       adsurl = {https://ui.adsabs.harvard.edu/abs/1991MNRAS.253..307P}
}

@ARTICLE{2017MNRAS,
       author = {{Wilson}, M.~J. and {Peacock}, J.~A. and {Taylor}, A.~N. and {de la Torre}, S.},
        title = "{Rapid modelling of the redshift-space power spectrum multipoles for a masked density field}",
         year = 2017,
        month = jan,
       volume = {464},
       number = {3},
        pages = {3121-3130},
          doi = {10.1093/mnras/stw2576},
archivePrefix = {arXiv},
       eprint = {1511.07799},
 primaryClass = {astro-ph.CO},
       adsurl = {https://ui.adsabs.harvard.edu/abs/2017MNRAS.464.3121W}
}

@article{Weaverdyck_2021,
   title="{Mitigating contamination in LSS surveys: a comparison of methods}",
   volume={503},
   ISSN={1365-2966},
   url={http://dx.doi.org/10.1093/mnras/stab709},
   DOI={10.1093/mnras/stab709},
   number={4},
   journal={Monthly Notices of the Royal Astronomical Society},
   publisher={Oxford University Press (OUP)},
   author={Weaverdyck, Noah and Huterer, Dragan},
   year={2021},
   month=Mar, pages={5061--5084} }

@article{Simon_2025,
   title="{Benchmarking field-level cosmological inference from galaxy redshift surveys}",
   volume={12},
   ISSN={1475-7516},
   url={http://dx.doi.org/10.1088/1475-7516/2025/12/039},
   DOI={10.1088/1475-7516/2025/12/039},
   number={12},
   journal={Journal of Cosmology and Astroparticle Physics},
   publisher={IOP Publishing},
   author={Simon, Hugo and Lanusse, François and de Mattia, Arnaud},
   year={2025},
   month=Dec, pages={039} }

@article{Schmidt_2019,
   title="{A rigorous EFT-based forward model for large-scale structure}",
   volume={01},
   ISSN={1475-7516},
   url={http://dx.doi.org/10.1088/1475-7516/2019/01/042},
   DOI={10.1088/1475-7516/2019/01/042},
   number={01},
   journal={Journal of Cosmology and Astroparticle Physics},
   publisher={IOP Publishing},
   author={Schmidt, Fabian and Elsner, Franz and Jasche, Jens and Nguyen, Nhat Minh and Lavaux, Guilhem},
   year={2019},
   month=Jan, pages={042} }

@article{Cabass_2020,
   title="{The EFT likelihood for large-scale structure}",
   volume={04},
   ISSN={1475-7516},
   url={http://dx.doi.org/10.1088/1475-7516/2020/04/042},
   DOI={10.1088/1475-7516/2020/04/042},
   number={04},
   journal={Journal of Cosmology and Astroparticle Physics},
   publisher={IOP Publishing},
   author={Cabass, Giovanni and Schmidt, Fabian},
   year={2020},
   month=Apr, pages={042} }

@article{Jasche_2013,
   title="{Bayesian physical reconstruction of initial conditions from large-scale structure surveys}",
   volume={432},
   ISSN={1365-2966},
   url={http://dx.doi.org/10.1093/mnras/stt449},
   DOI={10.1093/mnras/stt449},
   number={2},
   journal={Monthly Notices of the Royal Astronomical Society},
   publisher={Oxford University Press (OUP)},
   author={Jasche, Jens and Wandelt, Benjamin D.},
   year={2013},
   month=Apr, pages={894--913} }

@article{parker2026fieldlevelbaryonacousticoscillation,
      title="{Field-Level Baryon Acoustic Oscillation Reconstruction of the DESI DR1 Luminous Red Galaxies with Linear Field Transformer (LiFT)}", 
      author={Liam Parker and Uroš Seljak and Adrian E. Bayer and Richard Feder and David Valcin},
      year={2026},
      eprint={2609.14862},
      archivePrefix={arXiv},
      primaryClass={astro-ph.CO},
      url={https://arxiv.org/abs/2609.14862}, 
}

@article{Alsing_2018,
   title="{Generalized massive optimal data compression}",
   volume={476},
   ISSN={1745-3933},
   url={http://dx.doi.org/10.1093/mnrasl/sly029},
   DOI={10.1093/mnrasl/sly029},
   number={1},
   journal={Monthly Notices of the Royal Astronomical Society: Letters},
   publisher={Oxford University Press (OUP)},
   author={Alsing, Justin and Wandelt, Benjamin},
   year={2018},
   month=Feb, pages={L60--L64} }

@book{mathai1992quadratic,
  author    = {Mathai, A. M. and Provost, S. B.},
  title     = "{Quadratic Forms in Random Variables: Theory and Applications}",
  publisher = {Marcel Dekker, Inc.},
  address   = {New York},
  year      = {1992},
  series    = {Statistics: Textbooks and Monographs},
  volume    = {126}
}

@article{Ivanov_2020,
   title="{Cosmological parameters from the BOSS galaxy power spectrum}",
   volume={05},
   ISSN={1475-7516},
   url={http://dx.doi.org/10.1088/1475-7516/2020/05/042},
   DOI={10.1088/1475-7516/2020/05/042},
   number={05},
   journal={Journal of Cosmology and Astroparticle Physics},
   publisher={IOP Publishing},
   author={Ivanov, Mikhail M. and Simonović, Marko and Zaldarriaga, Matias},
   year={2020},
   month=May, pages={042} }

@article{Adame_2025,
   title="{DESI 2024 V: Full-Shape galaxy clustering from galaxies and quasars}",
   volume={09},
   ISSN={1475-7516},
   url={http://dx.doi.org/10.1088/1475-7516/2025/09/008},
   DOI={10.1088/1475-7516/2025/09/008},
   number={09},
   journal={Journal of Cosmology and Astroparticle Physics},
   publisher={IOP Publishing},
   author={Adame, A.G. and Aguilar, J. and Ahlen, S. and Alam, S. and Alexander, D.M. and Alvarez, M. and Alves, O. and Anand, A. and Andrade, U. and Armengaud, E. and Avila, S. and Aviles, A. and Awan, H. and Bailey, S. and Baltay, C. and Bault, A. and Behera, J. and BenZvi, S. and Beutler, F. and Bianchi, D. and Blake, C. and Blum, R. and Brieden, S. and Brodzeller, A. and Brooks, D. and Buckley-Geer, E. and Burtin, E. and Calderon, R. and Canning, R. and Carnero Rosell, A. and Cereskaite, R. and Cervantes-Cota, J.L. and Chabanier, S. and Chaussidon, E. and Chaves-Montero, J. and Chen, S. and Chen, X. and Claybaugh, T. and Cole, S. and Cuceu, A. and Davis, T.M. and Dawson, K. and de la Macorra, A. and de Mattia, A. and Deiosso, N. and Dey, A. and Dey, B. and Ding, Z. and Doel, P. and Edelstein, J. and Eftekharzadeh, S. and Eisenstein, D.J. and Elliott, A. and Fagrelius, P. and Fanning, K. and Ferraro, S. and Ereza, J. and Findlay, N. and Flaugher, B. and Font-Ribera, A. and Forero-Sánchez, D. and Forero-Romero, J.E. and Garcia-Quintero, C. and Garrison, L.H. and Gaztañaga, E. and Gil-Marín, H. and Gontcho, S.Gontcho A. and Gonzalez-Morales, A.X. and Gonzalez-Perez, V. and Gordon, C. and Green, D. and Gruen, D. and Gsponer, R. and Gutierrez, G. and Guy, J. and Hadzhiyska, B. and Hahn, C. and Hanif, M.M.S. and Herrera-Alcantar, H.K. and Honscheid, K. and Howlett, C. and Huterer, D. and Iršič, V. and Ishak, M. and Juneau, S. and Karaçaylı, N.G. and Kehoe, R. and Kent, S. and Kirkby, D. and Kong, H. and Koposov, S.E. and Kremin, A. and Krolewski, A. and Lai, Y. and Lan, T.-W. and Landriau, M. and Lang, D. and Lasker, J. and Le Goff, J.M. and Le Guillou, L. and Leauthaud, A. and Levi, M.E. and Li, T.S. and Lodha, K. and Magneville, C. and Manera, M. and Margala, D. and Martini, P. and Maus, M. and McDonald, P. and Medina-Varela, L. and Meisner, A. and Mena-Fernández, J. and Miquel, R. and Moon, J. and Moore, S. and Moustakas, J. and Mueller, E. and Muñoz-Gutiérrez, A. and Myers, A.D. and Nadathur, S. and Napolitano, L. and Neveux, R. and Newman, J.A. and Nguyen, N.M. and Nie, J. and Niz, G. and Noriega, H.E. and Padmanabhan, N. and Paillas, E. and Palanque-Delabrouille, N. and Pan, J. and Penmetsa, S. and Percival, W.J. and Pieri, M.M. and Pinon, M. and Poppett, C. and Porredon, A. and Prada, F. and Pérez-Fernández, A. and Pérez-Ràfols, I. and Rabinowitz, D. and Raichoor, A. and Ramírez-Pérez, C. and Ramirez-Solano, S. and Rashkovetskyi, M. and Ravoux, C. and Rezaie, M. and Rich, J. and Rocher, A. and Rockosi, C. and Rodríguez-Martínez, F. and Roe, N.A. and Rosado-Marin, A. and Ross, A.J. and Rossi, G. and Ruggeri, R. and Ruhlmann-Kleider, V. and Samushia, L. and Sanchez, E. and Saulder, C. and Schlafly, E.F. and Schlegel, D. and Schubnell, M. and Seo, H. and Sharples, R. and Silber, J. and Slosar, A. and Smith, A. and Sprayberry, D. and Tan, T. and Tarlé, G. and Trusov, S. and Vaisakh, R. and Valcin, D. and Valdes, F. and Vargas-Magaña, M. and Verde, L. and Walther, M. and Wang, B. and Wang, M.S. and Weaver, B.A. and Weaverdyck, N. and Wechsler, R.H. and Weinberg, D.H. and White, M. and Wilson, M.J. and Yu, J. and Yu, Y. and Yuan, S. and Yèche, C. and Zaborowski, E.A. and Zarrouk, P. and Zhang, H. and Zhao, C. and Zhao, R. and Zhou, R. and Zou, H.},
   year={2025},
   month=sep, pages={008} }

@article{Kosti_2023,
   title="{Consistency tests of field level inference with the EFT likelihood}",
   volume={07},
   ISSN={1475-7516},
   url={http://dx.doi.org/10.1088/1475-7516/2023/07/063},
   DOI={10.1088/1475-7516/2023/07/063},
   number={07},
   journal={Journal of Cosmology and Astroparticle Physics},
   publisher={IOP Publishing},
   author={Kostić, Andrija and Nguyen, Nhat-Minh and Schmidt, Fabian and Reinecke, Martin},
   year={2023},
   month=jul, pages={063} }

@article{leclercq2015probabilisticcartographylargescalestructure,
      title="{Probabilistic cartography of the large-scale structure}", 
      author={Florent Leclercq and Jens Jasche and Guilhem Lavaux and Benjamin Wandelt},
      year={2015},
      eprint={1512.02242},
      archivePrefix={arXiv},
      primaryClass={astro-ph.CO},
      url={https://arxiv.org/abs/1512.02242}, 
}

@article{Dalal_2008,
   title="{Imprints of primordial non-Gaussianities on large-scale structure: Scale-dependent bias and abundance of virialized objects}",
   volume={77},
   ISSN={1550-2368},
   url={http://dx.doi.org/10.1103/PhysRevD.77.123514},
   DOI={10.1103/physrevd.77.123514},
   number={12},
   journal={Physical Review D},
   publisher={American Physical Society (APS)},
   author={Dalal, Neal and Doré, Olivier and Huterer, Dragan and Shirokov, Alexander},
   year={2008},
   month=jun, pages={123514} }

@article{Desjacques_2010,
   title="{Primordial non-Gaussianity from the large-scale structure}",
   volume={27},
   ISSN={1361-6382},
   url={http://dx.doi.org/10.1088/0264-9381/27/12/124011},
   DOI={10.1088/0264-9381/27/12/124011},
   number={12},
   journal={Classical and Quantum Gravity},
   publisher={IOP Publishing},
   author={Desjacques, V and Seljak, U},
   year={2010},
   month=May, pages={124011} }

@article{Li_2019,
   title="{Disconnected covariance of 2-point functions in large-scale structure}",
   volume={01},
   ISSN={1475-7516},
   url={http://dx.doi.org/10.1088/1475-7516/2019/01/016},
   DOI={10.1088/1475-7516/2019/01/016},
   number={01},
   journal={Journal of Cosmology and Astroparticle Physics},
   publisher={IOP Publishing},
   author={Li, Yin and Singh, Sukhdeep and Yu, Byeonghee and Feng, Yu and Seljak, Uroš},
   year={2019},
   month=Jan, pages={016} }

@article{Yamamoto_2003,
   title="{Optimal Weighting Scheme in Redshift‐Space Power Spectrum Analysis and a Prospect for Measuring the Cosmic Equation of State}",
   volume={595},
   ISSN={1538-4357},
   url={http://dx.doi.org/10.1086/377488},
   DOI={10.1086/377488},
   number={2},
   journal={The Astrophysical Journal},
   publisher={American Astronomical Society},
   author={Yamamoto, Kazuhiro},
   year={2003},
   month=Oct, pages={577--588} }

@article{sailer2026optimalexactwideanglepower,
      title="{Optimal and exact wide-angle power spectrum estimation}", 
      author={Noah Sailer and Kendrick Smith and Yurii Kvasiuk and Alex Laguë and Selim Hotinli},
      year={2026},
      eprint={2607.05513},
      archivePrefix={arXiv},
      primaryClass={astro-ph.CO},
      url={https://arxiv.org/abs/2607.05513}, 
}

@article{Philcox_2021,
   title="{Cosmology without window functions: Quadratic estimators for the galaxy power spectrum}",
   volume={103},
   ISSN={2470-0029},
   url={http://dx.doi.org/10.1103/PhysRevD.103.103504},
   DOI={10.1103/physrevd.103.103504},
   number={10},
   journal={Physical Review D},
   publisher={American Physical Society (APS)},
   author={Philcox, Oliver H.E.},
   year={2021},
   month=May, pages={103504} }

@article{Hamilton_1997,
   title="{Towards optimal measurement of power spectra - I. Minimum variance pair weighting and the Fisher matrix}",
   volume={289},
   ISSN={1365-2966},
   url={http://dx.doi.org/10.1093/mnras/289.2.285},
   DOI={10.1093/mnras/289.2.285},
   number={2},
   journal={Monthly Notices of the Royal Astronomical Society},
   publisher={Oxford University Press (OUP)},
   author={Hamilton, A. J. S.},
   year={1997},
   month=Aug, pages={285--294} }

@article{Tegmark_1998,
   title="{Measuring the Galaxy Power Spectrum with Future Redshift Surveys}",
   volume={499},
   ISSN={1538-4357},
   url={http://dx.doi.org/10.1086/305663},
   DOI={10.1086/305663},
   number={2},
   journal={The Astrophysical Journal},
   publisher={American Astronomical Society},
   author={Tegmark, Max and Hamilton, Andrew J. S. and Strauss, Michael A. and Vogeley, Michael S. and Szalay, Alexander S.},
   year={1998},
   month=jun, pages={555--576} }

@article{fondi2026assemblybiaslocalprimordial,
      title="{Assembly bias and local Primordial non-Gaussianity from DESI DR1 Quasars}", 
      author={E. Fondi and L. Verde and E. Chaussidon and J. Aguilar and S. Ahlen and S. BenZvi and D. Bianchi and D. Brooks and T. Claybaugh and A. Cuceu and A. de la Macorra and P. Doel and S. Ferraro and J. E. Forero-Romero and E. Gaztañaga and S. Gontcho A Gontcho and G. Gutierrez and H. K. Herrera-Alcantar and D. Huterer and M. Ishak and R. Joyce and A. Kremin and O. Lahav and C. Lamman and M. Landriau and L. Le Guillou and M. Manera and P. Martini and A. Meisner and R. Miquel and S. Nadathur and N. Palanque-Delabrouille and W. J. Percival and F. Prada and I. Pérez-Ràfols and G. Rossi and L. Samushia and E. Sanchez and D. Schlegel and D. Sprayberry and G. Tarlé and B. A. Weaver and H. Zou},
      year={2026},
      eprint={2602.12357},
      archivePrefix={arXiv},
      primaryClass={astro-ph.CO},
      journal={Journal of Cosmology and Astroparticle Physics},
      volume={06},
      pages={062},
      doi={10.1088/1475-7516/2026/06/062},
      url={https://arxiv.org/abs/2602.12357}, 
}

@article{Castorina_2019_OQE,
   title="{Redshift-weighted constraints on primordial non-Gaussianity from the clustering of the eBOSS DR14 quasars in Fourier space}",
   volume={09},
   ISSN={1475-7516},
   url={http://dx.doi.org/10.1088/1475-7516/2019/09/010},
   DOI={10.1088/1475-7516/2019/09/010},
   number={09},
   journal={Journal of Cosmology and Astroparticle Physics},
   publisher={IOP Publishing},
   author={Castorina, Emanuele and Hand, Nick and Seljak, Uroš and Beutler, Florian and Chuang, Chia-Hsun and Zhao, Cheng and Gil-Marín, Héctor and Percival, Will J. and Ross, Ashley J. and Choi, Peter Doohyun and Dawson, Kyle and Macorra, Axel de la and Rossi, Graziano and Ruggeri, Rossana and Schneider, Donald and Zhao, Gong-Bo},
   year={2019},
   month=sep, pages={010} }

@article{Abdul_Karim_2025,
   title="{DESI DR2 results. II. Measurements of baryon acoustic oscillations and cosmological constraints}",
   volume={112},
   ISSN={2470-0029},
   url={http://dx.doi.org/10.1103/tr6y-kpc6},
   DOI={10.1103/tr6y-kpc6},
   number={8},
   journal={Physical Review D},
   publisher={American Physical Society (APS)},
   author={Abdul Karim, M. and Aguilar, J. and Ahlen, S. and Alam, S. and Allen, L. and Prieto, C. Allende and Alves, O. and Anand, A. and Andrade, U. and Armengaud, E. and Aviles, A. and Bailey, S. and Baltay, C. and Bansal, P. and Bault, A. and Behera, J. and BenZvi, S. and Bianchi, D. and Blake, C. and Brieden, S. and Brodzeller, A. and Brooks, D. and Buckley-Geer, E. and Burtin, E. and Calderon, R. and Canning, R. and Rosell, A. Carnero and Carrilho, P. and Casas, L. and Castander, F.J. and Charles, M. and Chaussidon, E. and Chaves-Montero, J. and Chebat, D. and Chen, X. and Claybaugh, T. and Cole, S. and Cooper, A.P. and Cuceu, A. and Dawson, K.S. and de la Macorra, A. and de Mattia, A. and Deiosso, N. and Della Costa, J. and Demina, R. and Dey, A. and Dey, B. and Ding, Z. and Doel, P. and Edelstein, J. and Eisenstein, D.J. and Elbers, W. and Fagrelius, P. and Fanning, K. and Fernández-García, E. and Ferraro, S. and Font-Ribera, A. and Forero-Romero, J.E. and Frenk, C.S. and Garcia-Quintero, C. and Garrison, L.H. and Gaztañaga, E. and Gil-Marín, H. and Gontcho, S. Gontcho A. and Gonzalez, D. and Gonzalez-Morales, A.X. and Gordon, C. and Green, D. and Gutierrez, G. and Guy, J. and Hadzhiyska, B. and Hahn, C. and He, S. and Herbold, M. and Herrera-Alcantar, H.K. and Ho, M.-F. and Honscheid, K. and Howlett, C. and Huterer, D. and Ishak, M. and Juneau, S. and Kamble, N.V. and Karaçayl, N.G. and Kehoe, R. and Kent, S. and Kim, A.G. and Kirkby, D. and Kisner, T. and Koposov, S.E. and Kremin, A. and Krolewski, A. and Lahav, O. and Lamman, C. and Landriau, M. and Lang, D. and Lasker, J. and Le Goff, J.M. and Le Guillou, L. and Leauthaud, A. and Levi, M.E. and Li, Q. and Li, T.S. and Lodha, K. and Lokken, M. and Lozano-Rodríguez, F. and Magneville, C. and Manera, M. and Martini, P. and Matthewson, W.L. and Meisner, A. and Mena-Fernández, J. and Menegas, A. and Mergulhão, T. and Miquel, R. and Moustakas, J. and Muñoz-Gutiérrez, A. and Muñoz-Santos, D. and Myers, A.D. and Nadathur, S. and Naidoo, K. and Napolitano, L. and Newman, J.A. and Niz, G. and Noriega, H.E. and Paillas, E. and Palanque-Delabrouille, N. and Pan, J. and Peacock, J.A. and Ibanez, M.P. and Percival, W.J. and Pérez-Fernández, A. and Pérez-Ràfols, I. and Pieri, M.M. and Poppett, C. and Prada, F. and Rabinowitz, D. and Raichoor, A. and Ramírez-Pérez, C. and Rashkovetskyi, M. and Ravoux, C. and Rich, J. and Rocher, A. and Rockosi, C. and Rohlf, J. and Román-Herrera, J.O. and Ross, A.J. and Rossi, G. and Ruggeri, R. and Ruhlmann-Kleider, V. and Samushia, L. and Sanchez, E. and Sanders, N. and Schlegel, D. and Schubnell, M. and Seo, H. and Shafieloo, A. and Sharples, R. and Silber, J. and Sinigaglia, F. and Sprayberry, D. and Tan, T. and Tarlé, G. and Taylor, P. and Turner, W. and Ureña-López, L.A. and Vaisakh, R. and Valdes, F. and Valogiannis, G. and Vargas-Magaña, M. and Verde, L. and Walther, M. and Weaver, B.A. and Weinberg, D.H. and White, M. and Wolfson, M. and Yèche, C. and Yu, J. and Zaborowski, E.A. and Zarrouk, P. and Zhai, Z. and Zhang, H. and Zhao, C. and Zhao, G.B. and Zhou, R. and Zou, H. },
   year={2025},
   month=Oct, pages={083515} }

@incollection{schmidt2006coping,
  author    = {Schmidt, Thorsten},
  title     = "{Coping with Copulas}",
  editor    = {Rank, J{\"o}rn},
  booktitle = "{Copulas: From Theory to Applications in Finance}",
  publisher = {Bloomberg Press},
  address   = {New York, NY},
  year      = {2006},
  month     = {dec},
  pages     = {1--31},
  isbn      = {978-1-904339-45-8}
}

@article{rosadomarín2026localprimordialnongaussianityusing,
      title="{Local primordial non-Gaussianity using cross-correlations of DESI tracers}", 
      author={A. J. Rosado-Marín and A. J. Ross and H. Seo and E. Chaussidon and J. Aguilar and S. Ahlen and D. Bianchi and D. Brooks and T. Claybaugh and A. Cuceu and A. de la Macorra and A. de Mattia and R. Demina and B. Dey and P. Doel and S. Ferraro and A. Font-Ribera and J. E. Forero-Romero and E. Gaztañaga and S. Gontcho A Gontcho and G. Gutierrez and C. Hahn and H. K. Herrera-Alcantar and D. Huterer and M. Ishak and R. Joyce and D. Kirkby and A. Kremin and O. Lahav and C. Lamman and M. Landriau and M. E. Levi and M. Manera and A. Meisner and R. Miquel and S. Nadathur and J. A. Newman and N. Palanque-Delabrouille and W. J. Percival and F. Prada and I. Pérez-Ràfols and G. Rossi and E. Sanchez and D. Schlegel and J. Silber and G. Tarlé and B. A. Weaver and C. Zhao and H. Zou},
      year={2026},
      eprint={2604.05213},
      archivePrefix={arXiv},
      primaryClass={astro-ph.CO},
      url={https://arxiv.org/abs/2604.05213}, 
}

@article{gorski1999healpixprimer,
      title="{The HEALPix Primer}", 
      author={Krzysztof M. Gorski and Benjamin D. Wandelt and Frode K. Hansen and Eric Hivon and Anthony J. Banday},
      year={1999},
      eprint={astro-ph/9905275},
      archivePrefix={arXiv},
      primaryClass={astro-ph},
      url={https://arxiv.org/abs/astro-ph/9905275}, 
}

@article{abdelaty1954approximate,
  author  = {Abdel-Aty, S. H.},
  title   = "{Approximate Formulae for the Percentage Points and the Probability Integral of the Non-Central {$\chi^2$} Distribution}",
  journal = {Biometrika},
  volume  = {41},
  number  = {3/4},
  pages   = {538--540},
  year    = {1954},
  month   = {dec},
  doi     = {10.1093/biomet/41.3-4.538},
  url     = {https://www.jstor.org/stable/2332731}
}

@article{2016_PLANCK_DUST,
   title="{Planck
                    intermediate results: XXX. The angular power spectrum of polarized dust emission at intermediate and high Galactic latitudes}",
   volume={586},
   ISSN={1432-0746},
   url={http://dx.doi.org/10.1051/0004-6361/201425034},
   DOI={10.1051/0004-6361/201425034},
   journal={Astronomy \& Astrophysics},
   publisher={EDP Sciences},
   collaboration={Planck},
   author={Adam, R. and Ade, P. A. R. and Aghanim, N. and Arnaud, M. and Aumont, J. and Baccigalupi, C. and Banday, A. J. and Barreiro, R. B. and Bartlett, J. G. and Bartolo, N. and Battaner, E. and Benabed, K. and Benoit-Lévy, A. and Bernard, J.-P. and Bersanelli, M. and Bielewicz, P. and Bonaldi, A. and Bonavera, L. and Bond, J. R. and Borrill, J. and Bouchet, F. R. and Boulanger, F. and Bracco, A. and Bucher, M. and Burigana, C. and Butler, R. C. and Calabrese, E. and Cardoso, J.-F. and Catalano, A. and Challinor, A. and Chamballu, A. and Chary, R.-R. and Chiang, H. C. and Christensen, P. R. and Clements, D. L. and Colombi, S. and Colombo, L. P. L. and Combet, C. and Couchot, F. and Coulais, A. and Crill, B. P. and Curto, A. and Cuttaia, F. and Danese, L. and Davies, R. D. and Davis, R. J. and de Bernardis, P. and de Zotti, G. and Delabrouille, J. and Delouis, J.-M. and Désert, F.-X. and Dickinson, C. and Diego, J. M. and Dolag, K. and Dole, H. and Donzelli, S. and Doré, O. and Douspis, M. and Ducout, A. and Dunkley, J. and Dupac, X. and Efstathiou, G. and Elsner, F. and Enßlin, T. A. and Eriksen, H. K. and Falgarone, E. and Finelli, F. and Forni, O. and Frailis, M. and Fraisse, A. A. and Franceschi, E. and Frejsel, A. and Galeotta, S. and Galli, S. and Ganga, K. and Ghosh, T. and Giard, M. and Giraud-Héraud, Y. and Gjerløw, E. and González-Nuevo, J. and Górski, K. M. and Gratton, S. and Gregorio, A. and Gruppuso, A. and Guillet, V. and Hansen, F. K. and Hanson, D. and Harrison, D. L. and Helou, G. and Henrot-Versillé, S. and Hernández-Monteagudo, C. and Herranz, D. and Hivon, E. and Hobson, M. and Holmes, W. A. and Huffenberger, K. M. and Hurier, G. and Jaffe, A. H. and Jaffe, T. R. and Jewell, J. and Jones, W. C. and Juvela, M. and Keihänen, E. and Keskitalo, R. and Kisner, T. S. and Kneissl, R. and Knoche, J. and Knox, L. and Krachmalnicoff, N. and Kunz, M. and Kurki-Suonio, H. and Lagache, G. and Lamarre, J.-M. and Lasenby, A. and Lattanzi, M. and Lawrence, C. R. and Leahy, J. P. and Leonardi, R. and Lesgourgues, J. and Levrier, F. and Liguori, M. and Lilje, P. B. and Linden-Vørnle, M. and López-Caniego, M. and Lubin, P. M. and Macías-Pérez, J. F. and Maffei, B. and Maino, D. and Mandolesi, N. and Mangilli, A. and Maris, M. and Martin, P. G. and Martínez-González, E. and Masi, S. and Matarrese, S. and Mazzotta, P. and Meinhold, P. R. and Melchiorri, A. and Mendes, L. and Mennella, A. and Migliaccio, M. and Mitra, S. and Miville-Deschênes, M.-A. and Moneti, A. and Montier, L. and Morgante, G. and Mortlock, D. and Moss, A. and Munshi, D. and Murphy, J. A. and Naselsky, P. and Nati, F. and Natoli, P. and Netterfield, C. B. and Nørgaard-Nielsen, H. U. and Noviello, F. and Novikov, D. and Novikov, I. and Pagano, L. and Pajot, F. and Paladini, R. and Paoletti, D. and Partridge, B. and Pasian, F. and Patanchon, G. and Pearson, T. J. and Perdereau, O. and Perotto, L. and Perrotta, F. and Pettorino, V. and Piacentini, F. and Piat, M. and Pierpaoli, E. and Pietrobon, D. and Plaszczynski, S. and Pointecouteau, E. and Polenta, G. and Ponthieu, N. and Popa, L. and Pratt, G. W. and Prunet, S. and Puget, J.-L. and Rachen, J. P. and Reach, W. T. and Rebolo, R. and Remazeilles, M. and Renault, C. and Renzi, A. and Ricciardi, S. and Ristorcelli, I. and Rocha, G. and Rosset, C. and Rossetti, M. and Roudier, G. and Rouillé d’Orfeuil, B. and Rubiño-Martín, J. A. and Rusholme, B. and Sandri, M. and Santos, D. and Savelainen, M. and Savini, G. and Scott, D. and Soler, J. D. and Spencer, L. D. and Stolyarov, V. and Stompor, R. and Sudiwala, R. and Sunyaev, R. and Sutton, D. and Suur-Uski, A.-S. and Sygnet, J.-F. and Tauber, J. A. and Terenzi, L. and Toffolatti, L. and Tomasi, M. and Tristram, M. and Tucci, M. and Tuovinen, J. and Valenziano, L. and Valiviita, J. and Van Tent, B. and Vibert, L. and Vielva, P. and Villa, F. and Wade, L. A. and Wandelt, B. D. and Watson, R. and Wehus, I. K. and White, M. and White, S. D. M. and Yvon, D. and Zacchei, A. and Zonca, A.},
   year={2016},
   month=Feb, pages={A133} }

@article{adolff2026itsenvironmentlocalfrm,
      title="{It's All About the Environment: Local $f_{\rm NL}$ from a Dark Matter Conditioned Multitracer Analysis}", 
      author={Julius Adolff and Uroš Seljak},
      year={2026},
      eprint={2609.04313},
      archivePrefix={arXiv},
      primaryClass={astro-ph.CO},
      url={https://arxiv.org/abs/2609.04313}, 
}

@article{Chudaykin_2025,
      title="{Reanalyzing DESI DR1: 3. Constraints on Inflation from Galaxy Power Spectra \& Bispectra}", 
      author={Anton Chudaykin and Mikhail M. Ivanov and Oliver H. E. Philcox},
      year={2026},
      eprint={2512.04266},
      archivePrefix={arXiv},
      primaryClass={astro-ph.CO},
      journal={Physical Review D},
      volume={113},
      pages={063552},
      doi={10.1103/fhj3-6q4x},
      url={https://arxiv.org/abs/2512.04266}, 
}

@article{Slosar_2008,
   title="{Constraints on local primordial non-Gaussianity from large scale structure}",
   volume={2008},
   ISSN={1475-7516},
   url={http://dx.doi.org/10.1088/1475-7516/2008/08/031},
   DOI={10.1088/1475-7516/2008/08/031},
   number={08},
   journal={Journal of Cosmology and Astroparticle Physics},
   publisher={IOP Publishing},
   author={Slosar, Anže and Hirata, Christopher and Seljak, Uroš and Ho, Shirley and Padmanabhan, Nikhil},
   year={2008},
   month=Aug, pages={031} }

@misc{desicollaboration2016desiexperimentisciencetargeting,
      title="{The DESI Experiment Part I: Science, Targeting, and Survey Design}", 
      author={{DESI Collaboration} and Amir Aghamousa and Jessica Aguilar and Steve Ahlen and Shadab Alam and Lori E. Allen and Carlos Allende Prieto and James Annis and Stephen Bailey and Christophe Balland and Otger Ballester and Charles Baltay and Lucas Beaufore and Chris Bebek and Timothy C. Beers and Eric F. Bell and José Luis Bernal and Robert Besuner and Florian Beutler and Chris Blake and Hannes Bleuler and Michael Blomqvist and Robert Blum and Adam S. Bolton and Cesar Briceno and David Brooks and Joel R. Brownstein and Elizabeth Buckley-Geer and Angela Burden and Etienne Burtin and Nicolas G. Busca and Robert N. Cahn and Yan-Chuan Cai and Laia Cardiel-Sas and Raymond G. Carlberg and Pierre-Henri Carton and Ricard Casas and Francisco J. Castander and Jorge L. Cervantes-Cota and Todd M. Claybaugh and Madeline Close and Carl T. Coker and Shaun Cole and Johan Comparat and Andrew P. Cooper and M. -C. Cousinou and Martin Crocce and Jean-Gabriel Cuby and Daniel P. Cunningham and Tamara M. Davis and Kyle S. Dawson and Axel de la Macorra and Juan De Vicente and Timothée Delubac and Mark Derwent and Arjun Dey and Govinda Dhungana and Zhejie Ding and Peter Doel and Yutong T. Duan and Anne Ealet and Jerry Edelstein and Sarah Eftekharzadeh and Daniel J. Eisenstein and Ann Elliott and Stéphanie Escoffier and Matthew Evatt and Parker Fagrelius and Xiaohui Fan and Kevin Fanning and Arya Farahi and Jay Farihi and Ginevra Favole and Yu Feng and Enrique Fernandez and Joseph R. Findlay and Douglas P. Finkbeiner and Michael J. Fitzpatrick and Brenna Flaugher and Samuel Flender and Andreu Font-Ribera and Jaime E. Forero-Romero and Pablo Fosalba and Carlos S. Frenk and Michele Fumagalli and Boris T. Gaensicke and Giuseppe Gallo and Juan Garcia-Bellido and Enrique Gaztanaga and Nicola Pietro Gentile Fusillo and Terry Gerard and Irena Gershkovich and Tommaso Giannantonio and Denis Gillet and Guillermo Gonzalez-de-Rivera and Violeta Gonzalez-Perez and Shelby Gott and Or Graur and Gaston Gutierrez and Julien Guy and Salman Habib and Henry Heetderks and Ian Heetderks and Katrin Heitmann and Wojciech A. Hellwing and David A. Herrera and Shirley Ho and Stephen Holland and Klaus Honscheid and Eric Huff and Timothy A. Hutchinson and Dragan Huterer and Ho Seong Hwang and Joseph Maria Illa Laguna and Yuzo Ishikawa and Dianna Jacobs and Niall Jeffrey and Patrick Jelinsky and Elise Jennings and Linhua Jiang and Jorge Jimenez and Jennifer Johnson and Richard Joyce and Eric Jullo and Stéphanie Juneau and Sami Kama and Armin Karcher and Sonia Karkar and Robert Kehoe and Noble Kennamer and Stephen Kent and Martin Kilbinger and Alex G. Kim and David Kirkby and Theodore Kisner and Ellie Kitanidis and Jean-Paul Kneib and Sergey Koposov and Eve Kovacs and Kazuya Koyama and Anthony Kremin and Richard Kron and Luzius Kronig and Andrea Kueter-Young and Cedric G. Lacey and Robin Lafever and Ofer Lahav and Andrew Lambert and Michael Lampton and Martin Landriau and Dustin Lang and Tod R. Lauer and Jean-Marc Le Goff and Laurent Le Guillou and Auguste Le Van Suu and Jae Hyeon Lee and Su-Jeong Lee and Daniela Leitner and Michael Lesser and Michael E. Levi and Benjamin L'Huillier and Baojiu Li and Ming Liang and Huan Lin and Eric Linder and Sarah R. Loebman and Zarija Lukić and Jun Ma and Niall MacCrann and Christophe Magneville and Laleh Makarem and Marc Manera and Christopher J. Manser and Robert Marshall and Paul Martini and Richard Massey and Thomas Matheson and Jeremy McCauley and Patrick McDonald and Ian D. McGreer and Aaron Meisner and Nigel Metcalfe and Timothy N. Miller and Ramon Miquel and John Moustakas and Adam Myers and Milind Naik and Jeffrey A. Newman and Robert C. Nichol and Andrina Nicola and Luiz Nicolati da Costa and Jundan Nie and Gustavo Niz and Peder Norberg and Brian Nord and Dara Norman and Peter Nugent and Thomas O'Brien and Minji Oh and Knut A. G. Olsen and Cristobal Padilla and Hamsa Padmanabhan and Nikhil Padmanabhan and Nathalie Palanque-Delabrouille and Antonella Palmese and Daniel Pappalardo and Isabelle Pâris and Changbom Park and Anna Patej and John A. Peacock and Hiranya V. Peiris and Xiyan Peng and Will J. Percival and Sandrine Perruchot and Matthew M. Pieri and Richard Pogge and Jennifer E. Pollack and Claire Poppett and Francisco Prada and Abhishek Prakash and Ronald G. Probst and David Rabinowitz and Anand Raichoor and Chang Hee Ree and Alexandre Refregier and Xavier Regal and Beth Reid and Kevin Reil and Mehdi Rezaie and Constance M. Rockosi and Natalie Roe and Samuel Ronayette and Aaron Roodman and Ashley J. Ross and Nicholas P. Ross and Graziano Rossi and Eduardo Rozo and Vanina Ruhlmann-Kleider and Eli S. Rykoff and Cristiano Sabiu and Lado Samushia and Eusebio Sanchez and Javier Sanchez and David J. Schlegel and Michael Schneider and Michael Schubnell and Aurélia Secroun and Uros Seljak and Hee-Jong Seo and Santiago Serrano and Arman Shafieloo and Huanyuan Shan and Ray Sharples and Michael J. Sholl and William V. Shourt and Joseph H. Silber and David R. Silva and Martin M. Sirk and Anze Slosar and Alex Smith and George F. Smoot and Debopam Som and Yong-Seon Song and David Sprayberry and Ryan Staten and Andy Stefanik and Gregory Tarle and Suk Sien Tie and Jeremy L. Tinker and Rita Tojeiro and Francisco Valdes and Octavio Valenzuela and Monica Valluri and Mariana Vargas-Magana and Licia Verde and Alistair R. Walker and Jiali Wang and Yuting Wang and Benjamin A. Weaver and Curtis Weaverdyck and Risa H. Wechsler and David H. Weinberg and Martin White and Qian Yang and Christophe Yeche and Tianmeng Zhang and Gong-Bo Zhao and Yi Zheng and Xu Zhou and Zhimin Zhou and Yaling Zhu and Hu Zou and Ying Zu},
      year={2016},
      eprint={1611.00036},
      archivePrefix={arXiv},
      primaryClass={astro-ph.IM},
      url={https://arxiv.org/abs/1611.00036}, 
}

@misc{robnik2025metropolisadjustedmicrocanonicalhamiltonian,
      title="{Metropolis Adjusted Microcanonical Hamiltonian Monte Carlo}", 
      author={Jakob Robnik and Reuben Cohn-Gordon and Uroš Seljak},
      year={2025},
      eprint={2503.01707},
      archivePrefix={arXiv},
      primaryClass={stat.CO},
      url={https://arxiv.org/abs/2503.01707}, 
}

\newpage
\appendix
\section{Optimality of the MUSE Estimator}\label{app:ProofOptimal}
In the main text, we argued that our field level analysis retains the information needed to attain the Cramér–Rao bound in Eq.~\eqref{eq:CramerRaoBound}. Here, we demonstrate this result explicitly. We begin with the gradient of the log-likelihood at the fiducial parameters $
    j_a = E_a - b_a
$,
where the estimator $E_a$ and bias term $b_a$ are defined as
\begin{equation}
    E_a = \frac{1}{2}\boldsymbol{d}_\Pi^\dagger  \mathbf C_\Pi^{-1}\frac{\partial \mathbf C_\Pi}{\partial \theta_a}\mathbf C_\Pi^{-1}\boldsymbol{d}_\Pi, \qquad b_a = \frac{1}{2}\mathrm{tr}\left(\mathbf C_\Pi^{-1}\frac{\partial \mathbf C_\Pi}{\partial \theta_a} \right).
\end{equation}
The Hessian of the log-likelihood, $\partial^2 \ell /\partial \theta_a \partial \theta_b$, depends on the particular data realization. In the main text, we instead used its ensemble average to define the Fisher matrix:
\begin{equation}
    F_{ab} = -\left\langle \frac{\partial^2 \ell}{\partial \theta_a \partial \theta_b} \right\rangle.
\end{equation}
With this substitution, our method is more precisely described as a Fisher scoring quadratic estimator than as a Laplace approximation, which typically uses the realization dependent Hessian to estimate the covariance. To evaluate the expectation, we again assume that ${\boldsymbol{\theta}}_{\rm fid}$ equals the true underlying parameters.

We compute the Hessian by differentiating $j_a$. Direct differentiation gives
\begin{align*}
    \frac{\partial E_a}{\partial \theta_b} &= -\frac{1}{2}\boldsymbol{d}_\Pi^\dagger  \mathbf C_\Pi^{-1}\frac{\partial \mathbf C_\Pi}{\partial \theta_b}\mathbf C_\Pi^{-1}\frac{\partial \mathbf C_\Pi}{\partial \theta_a}\mathbf C_\Pi^{-1}\boldsymbol{d}_\Pi \\ &\quad
    +\frac{1}{2}\boldsymbol{d}_\Pi^\dagger  \mathbf C_\Pi^{-1}\frac{\partial^2 \mathbf C_\Pi}{\partial \theta_a\partial \theta_b}\mathbf C_\Pi^{-1}\boldsymbol{d}_\Pi \\
    &\quad -\frac{1}{2}\boldsymbol{d}_\Pi^\dagger  \mathbf C_\Pi^{-1}\frac{\partial \mathbf C_\Pi}{\partial \theta_a}\mathbf C_\Pi^{-1}\frac{\partial \mathbf C_\Pi}{\partial \theta_b}\mathbf C_\Pi^{-1}\boldsymbol{d}_\Pi, \\
    \frac{\partial b_a}{\partial \theta_b} &= -\frac{1}{2}\mathrm{tr}\left(\mathbf C_\Pi^{-1}\frac{\partial \mathbf C_\Pi}{\partial \theta_b}\mathbf C_\Pi^{-1}\frac{\partial \mathbf C_\Pi}{\partial \theta_a} \right) \\
    &\quad +\frac{1}{2}\mathrm{tr}\left(\mathbf C_\Pi^{-1}\frac{\partial^2 \mathbf C_\Pi}{\partial \theta_a\partial \theta_b} \right).
\end{align*}
Here, ${\partial E_a}/{\partial \theta_b}$ describes how the estimator changes with the parameters assumed during inference. It should not be confused with the response matrix derivative used in the main text to estimate $F_{ab}$. Taking the expectation of $\partial E_a/\partial \theta_b$ and using the identity
\begin{equation}
    \langle \boldsymbol{d}_\Pi^\dagger \mathbf A\boldsymbol{d}_\Pi \rangle = \mathrm{tr}(\mathbf A\mathbf C_\Pi),
\end{equation}
which holds for any matrix $\mathbf A$, we obtain
\begin{equation}
    \left\langle \frac{\partial E_a}{\partial \theta_b} \right\rangle = -\mathrm{tr}\left(\mathbf C_\Pi^{-1}\frac{\partial \mathbf C_\Pi}{\partial \theta_a}\mathbf C_\Pi^{-1}\frac{\partial \mathbf C_\Pi}{\partial \theta_b} \right) + \frac{1}{2}\mathrm{tr}\left(\mathbf C_\Pi^{-1}\frac{\partial^2 \mathbf C_\Pi}{\partial \theta_a\partial \theta_b} \right).
\end{equation}
It follows that
\begin{equation}
    \left\langle \frac{\partial^2\ell}{\partial \theta_a \partial \theta_b} \right\rangle = \left\langle \frac{\partial E_a}{\partial \theta_b} - \frac{\partial b_a}{\partial \theta_b} \right\rangle = -\frac{1}{2}\mathrm{tr}\left(\mathbf C_\Pi^{-1}\frac{\partial \mathbf C_\Pi}{\partial \theta_a}\mathbf C_\Pi^{-1}\frac{\partial \mathbf C_\Pi}{\partial \theta_b} \right),
\end{equation}
as required.

\noindent We can also show directly, without relying on the accuracy of the quadratic expansion of $\ell$, that
\begin{equation}
    \mathrm{Cov}(\Delta \theta_a, \Delta \theta_b) = (F^{-1})_{ab}.
\end{equation}
Because $\mathbf F\Delta\boldsymbol{\theta} = \boldsymbol{j}$, this result follows if
\begin{equation}
    \mathrm{Cov}(j_a, j_b) = \mathrm{Cov}(E_a, E_b) = F_{ab}.
\end{equation}
For any symmetric matrices $\mathbf A$ and $\mathbf B$, the covariance of Gaussian quadratic forms is
\begin{equation}
    \operatorname{Cov}\left(
\boldsymbol d_\Pi^\dagger\mathbf A\boldsymbol d_\Pi,\,
\boldsymbol d_\Pi^\dagger\mathbf B\boldsymbol d_\Pi
\right)
=
2\operatorname{tr}
(\mathbf A\mathbf C_\Pi\mathbf B\mathbf C_\Pi).
\end{equation}
Substituting
\begin{equation}
    \mathbf A = \frac{1}{2}  \mathbf C_\Pi^{-1}({\partial \mathbf C_\Pi}/{\partial \theta_a})\mathbf C_\Pi^{-1}, \qquad \mathbf B = \frac{1}{2}  \mathbf C_\Pi^{-1}({\partial \mathbf C_\Pi}/{\partial \theta_b})\mathbf C_\Pi^{-1}
\end{equation}
gives
\begin{equation}
    \mathrm{Cov}(E_a, E_b) =\frac{1}{2}\mathrm{tr}\left(\mathbf C_\Pi^{-1}\frac{\partial \mathbf C_\Pi}{\partial \theta_a}\mathbf C_\Pi^{-1}\frac{\partial \mathbf C_\Pi}{\partial \theta_b} \right) = F_{ab},
\end{equation}
which establishes the claimed result.

\section{Illustration of the non-Gaussian Likelihood}
\label{App:Examples}

Here, we illustrate that the non-Gaussian likelihood~\eqref{eq:LogLikeSum} accurately captures the full field level parameter posterior while being significantly cheaper to evaluate. We consider two toy examples.

\begin{figure}
    \centering
    \includegraphics[width=0.75\linewidth]{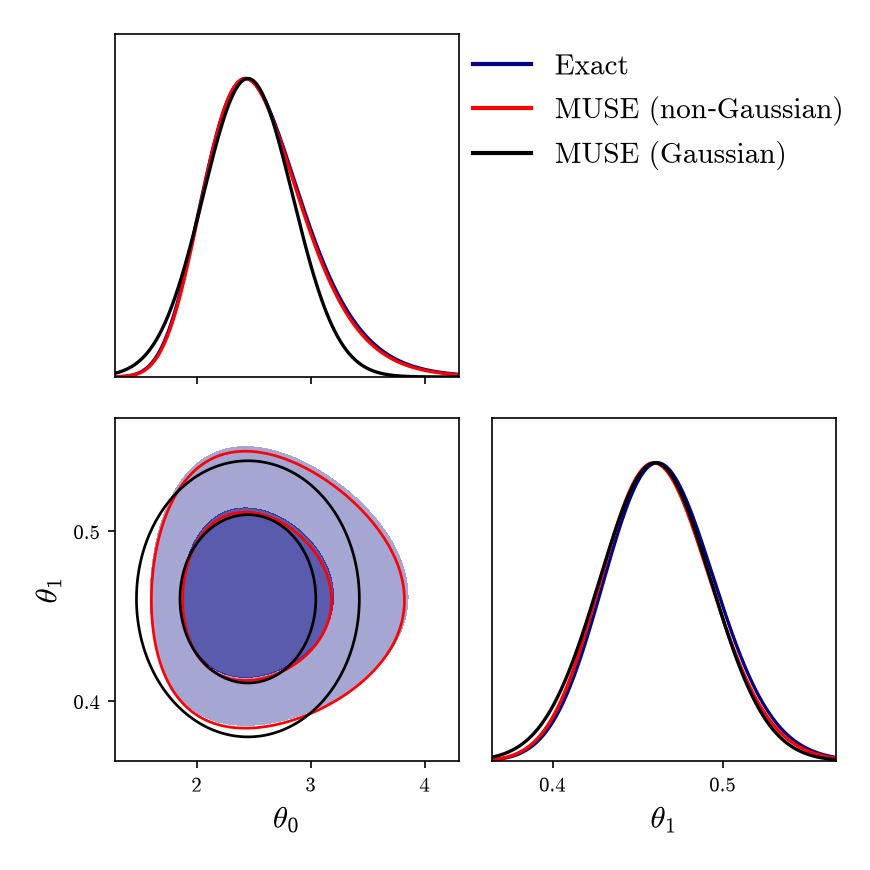}
    \caption{Comparison of the Gaussian and non-Gaussian approximations to the parameter posterior obtained with MUSE. The non-Gaussian corrections of Eq.~\eqref{eq:LogLikeSum} accurately capture the skewness of the exact posterior, which is known analytically to follow a Wishart distribution in this toy example.}
    \label{fig:Toy2}
\end{figure}

In Fig.~\ref{fig:Toy2}, we show the posterior of power spectrum broadband parameters in a simplified setting with constant noise, i.e. no window, for which the exact posterior is known analytically. Specifically, we take the data $\boldsymbol{d}$ to be a Gaussian field observed in a periodic box with uniform noise $\mathbf N = \sigma^2 \mathbf 1$. The forward model is the identity, and the prior is parameterized by the broadband parameters $\boldsymbol{\theta}$,
\begin{equation}
    \mathbf{P}(\boldsymbol{\theta}) = \sum_i \theta_i\,\mathbf \Pi_i,
\end{equation}
where $\mathbf \Pi_i$ projects onto the corresponding $k$ bin. In this case, the analytic distribution of the parameters $\boldsymbol{\theta}$ is known. They follow a Wishart distribution. In Fig.~\ref{fig:Toy2}, we compare this distribution with the posteriors obtained using MUSE and the approximation~\eqref{eq:LogLikeSum}.\footnote{To make the data linear in the unknown parameters, we sample $\sqrt{\theta_a}$ as part of the forward model using MUSE and analytically convert the resulting posterior into a posterior for $\boldsymbol{\theta}$.} We use a relatively low-dimensional setting, $\dim \boldsymbol{d} = 32^3$, and sample three broadband parameters so that their posteriors exhibit visible non-Gaussianity. We show the two lowest $k$ bins to illustrate how $p(\boldsymbol{E}\,|\,\boldsymbol{\theta})$ captures these features.

\begin{figure}
    \centering
    \includegraphics[width=1\linewidth]{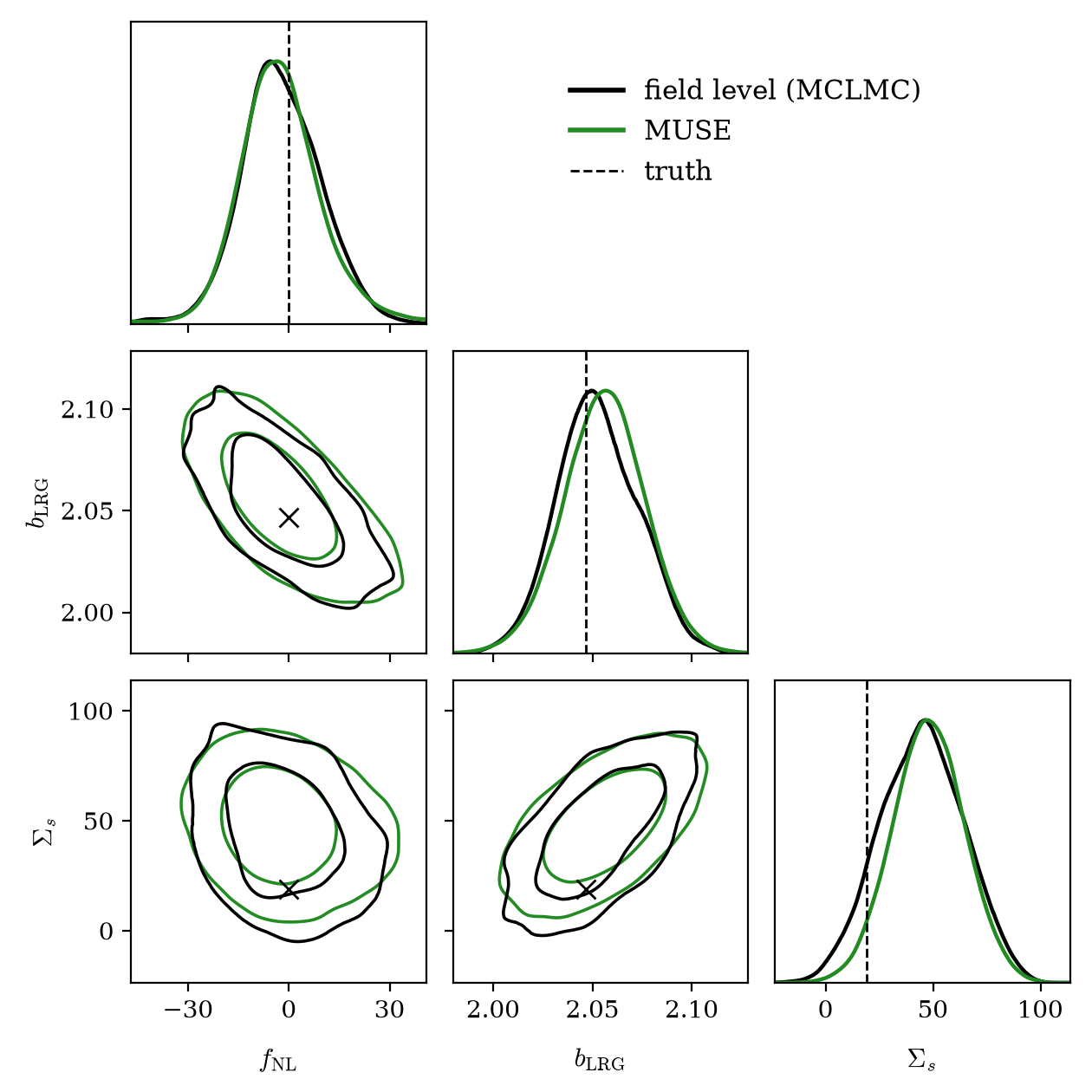}
    \caption{Comparison of the MUSE approximation~\eqref{eq:LogLikeSum} with a full field level analysis of the posterior, sampled using Microcanonical Langevin Monte Carlo (MCLMC)~\cite{robnik2025metropolisadjustedmicrocanonicalhamiltonian}. The two methods agree within sampling noise, which is relatively high here, with 250 effective samples.}
    \label{fig:Toy1}
\end{figure}

We also consider a more realistic cosmological setting by performing a full field level analysis of a simulated galaxy field. To keep the computational cost of sampling the $\sim 10^6$ dimensional latent space manageable, we consider only the LRG sample in the NGC and restrict the analysis to scales $k < k_{\rm max} = 0.05\,h/\mathrm{Mpc}$. This allows us to sample the initial field on a $128^3$ grid. For the field level analysis, we sampled the log-likelihood
\begin{equation}
    p(\boldsymbol{d},\boldsymbol{\delta}|\,\boldsymbol{\theta}) = -\frac{1}{2}(\boldsymbol{d}_\Pi - \boldsymbol{f}(\boldsymbol{\delta}, \boldsymbol{\theta}))^\dagger \mathbf N_\Pi^{-1}(\boldsymbol{d}_\Pi - \boldsymbol{f}(\boldsymbol{\delta}, \boldsymbol{\theta})) - \frac{1}{2}\boldsymbol{\delta}^\dagger \mathbf P^{-1}\boldsymbol{\delta} + \mathrm{const}
\end{equation}
and explicitly marginalize over $\boldsymbol{\delta}$ to obtain the parameter posterior $p(\boldsymbol{\theta}\,|\,\boldsymbol{d})$.
Even using MCLMC, which has been shown to be the most efficient sampler in the cosmological setting~\cite{Simon_2025}, inference for a single mock realization required more than $6$ GPU hours when initialized at the MUSE posterior maximum, yielding an effective sample size of about $250$ for $f_{\rm NL}$. On the $256^3$ grid required for the main analysis, this would scale to several hundred GPU hours. In comparison, calibrating the MUSE likelihood~\eqref{eq:LogLikeSum} required less than $0.2$~GPU hours. Moreover, once MUSE is calibrated, applying it to additional mocks is essentially free. Generating coverage plots, such as those shown in the main text, is therefore inexpensive with MUSE but prohibitively so when sampling the latent field separately for every mock. This is a major advantage of the present method over explicit field level inference. Figure~\ref{fig:Toy1} shows that the two methods agree up to sampling noise.

\section{Impact of the Integral Constraints }\label{app:IC}
As we have shown, the impact of the radial and angular integral constraints can be understood as applying the projector $\mathbf{\Pi}_{\rm RIC}\mathbf \Pi_{\rm AIC}$ to the FKP data vector $\boldsymbol{d}$. From this we can compute the impact of the integral constraints on the measured power spectrum exactly. Conceptually, this can be understood as a correction to the window matrix and to the shot noise subtraction.

\subsection{The Projector Acting on the Data}
We begin by writing both integral constraints in one common form. The Poisson noise matrix of $\boldsymbol d$ is $\mathbf N =c\,\mathbf W$, where $\mathbf W$ is the window operator of Sec.~\ref{sec:Window} and
\begin{equation}
    c \equiv \langle w_g^2\rangle+\alpha\langle w_r^2\rangle,
\end{equation}
where $w_g$ and $w_r$ are the completeness weights of the galaxies and randoms. A constant rescaling of the metric leaves a projector unchanged. For the angular integral constraint, Sec.~\ref{sec:IC} introduced the basis $\{\boldsymbol e_i\}$ of the span of the templates $\{\mathbf W \boldsymbol t_i\}$. We normalize it as $\boldsymbol e_i^\dagger \mathbf W^{-1}\boldsymbol e_j = \delta_{ij}$, so that $\mathbf \Pi_{\rm AIC} = \boldsymbol 1 - \sum_i \boldsymbol e_i\boldsymbol e_i^\dagger \mathbf W^{-1}$. The radial integral constraint takes the same form. With $\boldsymbol{r}_i = \mathbf 1_{[z_i, z_i + \Delta z_i]}$ the indicator of the $i$th shell, we define
\begin{equation}
    \boldsymbol e_i = \frac{\mathbf W \boldsymbol r_i}{\sqrt{\boldsymbol r_i^\dagger \mathbf W\boldsymbol r_i}}.
\end{equation}
These vectors also satisfy $\boldsymbol e_i^\dagger \mathbf W^{-1}\boldsymbol e_j = \delta_{ij}$ because the shells are disjoint, and 
\begin{equation}
    \boldsymbol e_i\boldsymbol e_i^\dagger \mathbf W^{-1} = \mathbf W \boldsymbol r_i\boldsymbol r_i^\dagger / (\boldsymbol r_i^\dagger \mathbf W\boldsymbol r_i).
\end{equation}
Hence $\mathbf \Pi_{\rm RIC} = \boldsymbol 1 - \sum_i \boldsymbol e_i\boldsymbol e_i^\dagger \mathbf W^{-1}$ as well. We project the templates with $\mathbf \Pi_{\rm RIC}$ before orthonormalizing them. The two sets of basis vectors are then mutually orthogonal, and the combined projector is
\begin{equation}
    \mathbf \Pi_{\rm IC} \equiv \mathbf{\Pi}_{\rm RIC}\mathbf \Pi_{\rm AIC} = \boldsymbol 1 - \sum_m \boldsymbol e_m\boldsymbol e_m^\dagger \mathbf W^{-1}, \qquad \boldsymbol e_m^\dagger \mathbf W^{-1}\boldsymbol e_n = \delta_{mn},
\end{equation}
where $m$ runs over the shells and the templates. It satisfies $\mathbf \Pi_{\rm IC}^2 = \mathbf \Pi_{\rm IC}$ and $\mathbf \Pi_{\rm IC}\boldsymbol e_m = 0$.

The weighted FKP field is formed from the deprojected data, $\boldsymbol F_{\rm IC} = \boldsymbol{w}\,\mathbf \Pi_{\rm IC}\boldsymbol{d}$, where $\boldsymbol w$ is the diagonal matrix of the FKP weights $w(\boldsymbol x)$. We write $\boldsymbol F_{\rm IC} = \mathbf \Pi_F\,\boldsymbol F$, where $\boldsymbol F = \boldsymbol w \boldsymbol d$, with the projector acting on the weighted field defined as
\begin{equation}
    \mathbf \Pi_F = \boldsymbol{w}\,\mathbf \Pi_{\rm IC}\,\boldsymbol{w}^{-1} = \boldsymbol 1 - \sum_m \boldsymbol v_m\boldsymbol w_m^\dagger, \label{eq:PiF}
\end{equation}
where
\begin{equation}
    \boldsymbol v_m = \boldsymbol w\,\boldsymbol e_m, \qquad \boldsymbol w_m = \boldsymbol w^{-1}\mathbf W^{-1}\boldsymbol e_m.
\end{equation}
We note for later use that $\boldsymbol w_m = (\boldsymbol w\mathbf W\boldsymbol w)^{-1}\boldsymbol v_m$. The matrix $\boldsymbol w\mathbf W\boldsymbol w$ is the Poisson noise matrix of the weighted field, up to the constant $c$. The $\boldsymbol v_m$ are therefore orthonormal with respect to $(\boldsymbol w\mathbf W\boldsymbol w)^{-1}$, $\boldsymbol w_m^\dagger \boldsymbol v_n = \delta_{mn}$, and $\mathbf \Pi_F$ is the orthogonal projector in the noise metric of $\boldsymbol F$. Below we write $\Pi_{\rm IC}(\boldsymbol{k}, \boldsymbol{q})$ for the matrix elements of $\mathbf \Pi_F$, that is,
\begin{align}
    \Pi_{\rm IC}(\boldsymbol{k}, \boldsymbol{q}) &= (2\pi)^3\,\delta_D(\boldsymbol{k} - \boldsymbol{q}) - \sum_m v_m(\boldsymbol{k})w_m^*(\boldsymbol{q}). \label{eq:PiModes}
\end{align}
The deprojected field can be written as
\begin{equation}
    \boldsymbol F_{\rm IC} =\boldsymbol F - \sum_m \boldsymbol{v}_m\,\langle \boldsymbol{w}_m, \boldsymbol F\rangle,
\end{equation}
with $\langle \boldsymbol{w}_m,\boldsymbol F\rangle = \int_{\boldsymbol{y}}w_m^*(\boldsymbol{y})F(\boldsymbol{y}) = \int_{\boldsymbol{q}}w_m^*(\boldsymbol{q})F(\boldsymbol{q})$.

\subsection{Two-Point Function of the Deprojected Field}
We assume the two-point function of the data to be of the local plane parallel form
\begin{equation}
    \langle d(\boldsymbol{x})\,d(\boldsymbol{y})\rangle = \bar n(\boldsymbol{x})\,\bar n(\boldsymbol{y})\sum_{\ell'}\int_{\boldsymbol{p}}P_{\ell'}(p)\,\mathcal L_{\ell'}(\hat{\boldsymbol{p}}\,\hat{\boldsymbol{x}})\,e^{i\boldsymbol{p}(\boldsymbol{x}-\boldsymbol{y})} + c\,\bar n(\boldsymbol{x})\,\delta_D(\boldsymbol{x}-\boldsymbol{y}), \label{eq:TheoryForm}
\end{equation}
which neglects wide angle corrections. The sum runs over the even multipoles of the theory, $\ell' = 0, 2, 4$ for the linear model used here. For the weighted field $F(\boldsymbol{x}) = w(\boldsymbol{x})\,d(\boldsymbol{x})$, we define the mask
\begin{equation}
    W(\boldsymbol{k}) = \int_{\boldsymbol{x}}e^{-i\boldsymbol{k}\boldsymbol{x}}\,\bar n(\boldsymbol{x})w(\boldsymbol{x}),
\end{equation}
and its Legendre-weighted generalization
\begin{equation}
    W_{\ell\ell'}(\boldsymbol{k}, \boldsymbol{p}) = \int_{\boldsymbol{x}}e^{-i(\boldsymbol{k}-\boldsymbol{p})\boldsymbol{x}}\,\bar n(\boldsymbol{x})w(\boldsymbol{x})\,\mathcal L_{\ell}(\hat{\boldsymbol{k}}\,\hat{\boldsymbol{x}})\,\mathcal L_{\ell'}(\hat{\boldsymbol{p}}\,\hat{\boldsymbol{x}}),
\end{equation}
so that $W_{00}(\boldsymbol{k}, \boldsymbol{p}) = W(\boldsymbol{k} - \boldsymbol{p})$. Additionally, we define the noise kernel as
\begin{equation}
    \mathcal S(\boldsymbol{k}) = \int_{\boldsymbol{x}} \bar n(\boldsymbol{x})w(\boldsymbol{x})^2e^{-i\boldsymbol{k}\boldsymbol{x}}.
\end{equation}
Denoting $\mathcal W_{\ell'}(\boldsymbol{k}, \boldsymbol{q}\,|\,\boldsymbol{p}) \equiv W_{0\ell'}(\boldsymbol{k}, \boldsymbol{p})\,W(\boldsymbol{p} - \boldsymbol{q})$, Eq.~\eqref{eq:TheoryForm} gives the standard FKP two-point function
\begin{equation}
    \langle F(\boldsymbol{k})F(\boldsymbol{q})^*\rangle = \sum_{\ell'}\int_{\boldsymbol{p}}\mathcal W_{\ell'}(\boldsymbol{k}, \boldsymbol{q}\,|\,\boldsymbol{p})P_{\ell'}(p) + c\,\mathcal S(\boldsymbol{k} - \boldsymbol{q}). \label{eq:StandardFKP}
\end{equation}
The deprojection acts on both fields of the two-point function. We find
\begin{equation}
    \langle F_{\rm IC}(\boldsymbol{k})F_{\rm IC}(\boldsymbol{q})^*\rangle = \int_{\boldsymbol{k}', \boldsymbol{q}'}\Pi_{\rm IC}(\boldsymbol{k}, \boldsymbol{k}')\,\Pi_{\rm IC}^*(\boldsymbol{q}, \boldsymbol{q}')\,\langle F(\boldsymbol{k}')F(\boldsymbol{q}')^*\rangle.
\end{equation}
With that, the correction of Eq.~\eqref{eq:StandardFKP} in the presence of the integral constraints can be written as the substitutions
\begin{equation}
    \mathcal W_{\ell'}(\boldsymbol{k}, \boldsymbol{q}\,|\,\boldsymbol{p}) \rightarrow \mathcal W_{{\rm IC},\ell'}(\boldsymbol{k}, \boldsymbol{q}\,|\,\boldsymbol{p}), \qquad \mathcal S(\boldsymbol{k} - \boldsymbol{q}) \rightarrow \mathcal S_{\rm IC}(\boldsymbol{k}, \boldsymbol{q}),
\end{equation}
where we defined
\begin{equation}
    \mathcal W_{{\rm IC},\ell'}(\boldsymbol{k}, \boldsymbol{q}\,|\,\boldsymbol{p}) = \int_{\boldsymbol{k}', \boldsymbol{q}'}\Pi_{\rm IC}(\boldsymbol{k}, \boldsymbol{k}')\Pi_{\rm IC}^*(\boldsymbol{q}, \boldsymbol{q}')\mathcal W_{\ell'}(\boldsymbol{k}', \boldsymbol{q}'\,|\,\boldsymbol{p})
\end{equation}
and
\begin{equation}
    \mathcal S_{\rm IC}(\boldsymbol{k}, \boldsymbol{q}) = \int_{\boldsymbol{k}', \boldsymbol{q}'}\Pi_{\rm IC}(\boldsymbol{k}, \boldsymbol{k}')\Pi_{\rm IC}^*(\boldsymbol{q}, \boldsymbol{q}')\,\mathcal S(\boldsymbol{k}' - \boldsymbol{q}').
\end{equation}
We now take the limit $\boldsymbol{k} \rightarrow \boldsymbol{q}$. The function $\mathcal W_{\ell'}(\boldsymbol{k}', \boldsymbol{q}'\,|\,\boldsymbol{p})$ factorizes into a function of $\boldsymbol{k}'$ and a function of $\boldsymbol{q}'$. The integral constraint corrected two-point function therefore takes the form
\begin{equation}
    \langle |F_{\rm IC}(\boldsymbol{k})|^2\rangle = \sum_{\ell'}\int_{\boldsymbol{p}} W_{{\rm IC},0\ell'} (\boldsymbol k, \boldsymbol{p})\,W_{{\rm IC},00} (\boldsymbol k, \boldsymbol{p})^*\,P_{\ell'}(p) + c\,S_{\rm IC}(\boldsymbol{k}), \label{eq:CorrectedFKP}
\end{equation}
where we defined
\begin{equation}
    W_{{\rm IC},0\ell'} (\boldsymbol k, \boldsymbol{p}) = \int_{\boldsymbol{k}'} \Pi_{\rm IC}(\boldsymbol{k}, \boldsymbol{k}')W_{0\ell'}(\boldsymbol{k}', \boldsymbol{p}), \label{eq:WIC}
\end{equation}
as well as
\begin{equation}
     S_{\rm IC}(\boldsymbol{k}) = \int_{\boldsymbol{k}', \boldsymbol{q}'}\Pi_{\rm IC}(\boldsymbol{k}, \boldsymbol{k}')\,\mathcal S(\boldsymbol{k}' - \boldsymbol{q}')\,\Pi_{\rm IC}^*(\boldsymbol{k}, \boldsymbol{q}'). \label{eq:SIC}
\end{equation}
For an isotropic theory Eq.~\eqref{eq:CorrectedFKP} reduces to $\int_{\boldsymbol{p}} |W_{{\rm IC},00} (\boldsymbol k, \boldsymbol{p})|^2 P(p)$. The window ${|W(\boldsymbol{k} - \boldsymbol{p})|^2}$ is thus replaced by the squared modulus of the deprojected window $W_{{\rm IC},00}(\boldsymbol k, \boldsymbol{p})$.

\subsection{Multipoles}
To derive the Yamamoto estimator in the current setup we discretize $k$ into bins $k_i$, with phase space volume $V_i$, and write
\begin{equation}
    \hat P_\ell(k_i) = \frac{2 \ell + 1}{V_{i}}\int_{V_{i}(\boldsymbol{k})} \mathcal F_\ell(\boldsymbol{k}) \mathcal F_0^*(\boldsymbol{k}) - \mathcal N_\ell(k_i),
\end{equation}
where, if $\hat{\boldsymbol{x}}$ is the local line of sight, we have defined
\begin{equation}
    \mathcal F_\ell(\boldsymbol{k}) = \int_{\boldsymbol{x}}e^{-i\boldsymbol{k}\boldsymbol{x}}\,F_{\rm IC}(\boldsymbol{x})\,\mathcal{L}_\ell(\hat{\boldsymbol{k}}\,\hat{\boldsymbol{x}}), \qquad \mathcal F_0(\boldsymbol{k}) = F_{\rm IC}(\boldsymbol{k}).
\end{equation}
The term $\mathcal N_\ell(k_i)$ is defined such that the theory multipole moments $P_{\ell'}(k_j)$ and the measurement $\hat P_\ell(k_i)$ are related linearly,
\begin{equation}
    \hat P_\ell(k_i) = \sum_{j, \ell '}(\mathcal{W}_{\ell \ell'})_{ij}\,P_{\ell'}(k_j),
\end{equation}
where the theory multipoles are taken to be constant within each bin $V_j$. To evaluate $\langle \mathcal F_\ell(\boldsymbol{k})\mathcal F_0^*(\boldsymbol{k})\rangle$, we insert $\boldsymbol F_{\rm IC} = \boldsymbol F - \sum_m \boldsymbol{v}_m\,\langle \boldsymbol{w}_m,\boldsymbol F\rangle$ into both factors. This produces four terms, each of which we evaluate with Eq.~\eqref{eq:TheoryForm}. The result is expressed through the Legendre-weighted transforms of the projected modes,
\begin{equation}
    v_m^{(\ell)}(\boldsymbol{k}) = \int_{\boldsymbol{x}}e^{-i\boldsymbol{k}\boldsymbol{x}}\,v_m(\boldsymbol{x})\,\mathcal L_\ell(\hat{\boldsymbol{k}}\,\hat{\boldsymbol{x}}), \label{eq:Images}
\end{equation}
through their overlaps with the mask and with the noise,
\begin{align}
    \tilde{w}_{m,\ell'}(\boldsymbol{p}) &= \int_{\boldsymbol{x}}e^{-i\boldsymbol{p}\boldsymbol{x}}\,\bar n(\boldsymbol{x})w(\boldsymbol{x})\,w_m(\boldsymbol{x})\,\mathcal L_{\ell'}(\hat{\boldsymbol{p}}\,\hat{\boldsymbol{x}}), \nonumber\\
    s_m^{(\ell)}(\boldsymbol{k}) &= \int_{\boldsymbol{x}}e^{-i\boldsymbol{k}\boldsymbol{x}}\,\bar n(\boldsymbol{x})w(\boldsymbol{x})^2\,w_m(\boldsymbol{x})\,\mathcal L_\ell(\hat{\boldsymbol{k}}\,\hat{\boldsymbol{x}}), \label{eq:Overlaps}
\end{align}
as well as the matrix 
\begin{equation}
    Q_{mn} = \int_{\boldsymbol{x}}\bar n(\boldsymbol{x})w(\boldsymbol{x})^2\,w_m^*(\boldsymbol{x})\,w_n(\boldsymbol{x}),
\end{equation}
and 
\begin{equation}
    \mathcal S_\ell(\boldsymbol{k}) = \int_{\boldsymbol{x}}\bar n(\boldsymbol{x})w(\boldsymbol{x})^2\,\mathcal L_\ell(\hat{\boldsymbol{k}}\,\hat{\boldsymbol{x}}),
\end{equation}
so that $\mathcal S_0 = \mathcal S(0)$. We write $v_m \equiv v_m^{(0)}$, $\tilde w_m \equiv \tilde w_{m,0}$ and $s_m \equiv s_m^{(0)}$. We note that ${\tilde w_m^*(\boldsymbol{p}) = \int_{\boldsymbol{q}}W(\boldsymbol{q} - \boldsymbol{p})w_m^*(\boldsymbol{q})}$ is the convolution of the mask with the projected mode. This is a product in real space, so each of these transforms costs one FFT. In terms of these quantities the two-point function of the multipoles takes the same form as Eq.~\eqref{eq:CorrectedFKP},
\begin{equation}
    \langle \mathcal F_\ell(\boldsymbol{k})\mathcal F_0^*(\boldsymbol{k})\rangle = \sum_{\ell'}\int_{\boldsymbol{p}} W_{{\rm IC},\ell\ell'} (\boldsymbol k, \boldsymbol{p})\,W_{{\rm IC},00} (\boldsymbol k, \boldsymbol{p})^*\,P_{\ell'}(p) + c\,S_{{\rm IC},\ell}(\boldsymbol{k}). \label{eq:CorrectedMultipoles}
\end{equation}
The deprojected window and shot noise are the usual ones plus corrections of rank $m$,
\begin{align}
    W_{{\rm IC},\ell\ell'} (\boldsymbol k, \boldsymbol{p}) &= W_{\ell\ell'}(\boldsymbol{k}, \boldsymbol{p}) - \sum_m v_m^{(\ell)}(\boldsymbol{k})\,\tilde{w}_{m,\ell'}^*(\boldsymbol{p}), \label{eq:WindowCorrected}\\
    S_{{\rm IC},\ell}(\boldsymbol{k}) &= \mathcal S_\ell(\boldsymbol{k}) - \sum_m v_m^{(\ell)}(\boldsymbol{k})\,s_m^*(\boldsymbol{k}) - \sum_n v_n^*(\boldsymbol{k})\,s_n^{(\ell)}(\boldsymbol{k}) \nonumber\\
    &\quad + \sum_{m,n}v_m^{(\ell)}(\boldsymbol{k})\,v_n^*(\boldsymbol{k})\,Q_{mn}. \label{eq:NoiseCorrected}
\end{align}
For $\ell = 0$, these are Eqs.~\eqref{eq:WIC} and~\eqref{eq:SIC} with the form~\eqref{eq:PiModes} of the projector inserted. In particular, $W_{{\rm IC},00} (\boldsymbol k, \boldsymbol{p}) = W(\boldsymbol{k} - \boldsymbol{p}) - \sum_m v_m(\boldsymbol{k})\tilde{w}_m^*(\boldsymbol{p})$ and
\begin{equation}
    S_{\rm IC}(\boldsymbol{k}) = \mathcal S_0 - 2\,\Re\sum_m v_m(\boldsymbol{k})s_m^*(\boldsymbol{k}) + \sum_{m,n}v_m(\boldsymbol{k})v_n^*(\boldsymbol{k})\,Q_{mn}.
\end{equation}
The projected modes thus reduce the shot noise by a $\boldsymbol{k}$-dependent amount. This correction is concentrated on the scales where the projected modes carry their power. The window matrix and the shot noise correction follow as
\begin{equation}
    (\mathcal{W}_{\ell\ell'})_{ij} = \frac{2\ell+1}{V_i}\int_{V_i(\boldsymbol{k})}\int_{V_j(\boldsymbol{p})}W_{{\rm IC},\ell\ell'} (\boldsymbol k, \boldsymbol{p})\,W_{{\rm IC},00} (\boldsymbol k, \boldsymbol{p})^*, \label{eq:WindowMatrixWrittenOut}
\end{equation}
and
\begin{equation}
    \mathcal N_\ell(k_i) = c\,\frac{2\ell+1}{V_i}\int_{V_i(\boldsymbol{k})} S_{{\rm IC},\ell}(\boldsymbol{k}). \label{eq:WindowAndNoise}
\end{equation}
For the window matrix it is useful to define
\begin{equation}
    A^{(\ell)}_{mn}(\boldsymbol{k}) = v^{(\ell)}_m(\boldsymbol{k})\,v_n^*(\boldsymbol{k}), \qquad B^{(\ell')}_{mn}(\boldsymbol{p}) = \tilde{w}_{m,\ell'}^*(\boldsymbol{p})\,\tilde{w}_n(\boldsymbol{p}),
\end{equation}
in terms of which Eq.~\eqref{eq:WindowCorrected} gives
\begin{align}
    W_{{\rm IC},\ell\ell'} (\boldsymbol k, \boldsymbol{p})\,W_{{\rm IC},00} (\boldsymbol k, \boldsymbol{p})^* &= W_{\ell\ell'}(\boldsymbol{k}, \boldsymbol{p})\,W(\boldsymbol{k} - \boldsymbol{p})^* - \sum_n v_n^*(\boldsymbol{k})\,W_{\ell\ell'}(\boldsymbol{k}, \boldsymbol{p})\,\tilde{w}_n(\boldsymbol{p}) \nonumber\\
    &\quad - \sum_m v_m^{(\ell)}(\boldsymbol{k})\,\tilde{w}_{m,\ell'}^*(\boldsymbol{p})\,W(\boldsymbol{k} - \boldsymbol{p})^* + \sum_{m,n} A^{(\ell)}_{mn}(\boldsymbol{k})\,B^{(\ell')}_{mn}(\boldsymbol{p}). \nonumber
\end{align}
The first term is the standard window. The last term is a sum of products of quantities that are precomputed once. The two cross terms are convolutions of the mask with a known function. Inserting Eq.~\eqref{eq:NoiseCorrected} and the above in Eq.~\eqref{eq:WindowAndNoise} and Eq.~\eqref{eq:WindowMatrixWrittenOut} gives the window matrix and the shot noise correction in closed form for any projector of the form~\eqref{eq:PiModes}.

\subsection{Simplification for the Noise Orthonormal Projector}
For the projector of Eq.~\eqref{eq:PiF}, the vectors $\boldsymbol w_m = (\boldsymbol w\mathbf W\boldsymbol w)^{-1}\boldsymbol v_m$ are tied to the images $\boldsymbol v_m$. The auxiliary quantities of Eq.~\eqref{eq:Overlaps} then collapse. Inserting $\bar n w^2\,w_m = v_m$ and $\bar n w\,w_m = e_m$, we find
\begin{equation}
    Q_{mn} = \delta_{mn}, \qquad s_m^{(\ell)}(\boldsymbol{k}) = v_m^{(\ell)}(\boldsymbol{k}), \qquad \tilde{w}_{m,\ell'}(\boldsymbol{p}) = e_m^{(\ell')}(\boldsymbol{p}),
\end{equation}
where we defined
\begin{equation}
    e_m^{(\ell')}(\boldsymbol{p}) = \int_{\boldsymbol{x}}e^{-i\boldsymbol{p}\boldsymbol{x}}\,e_m(\boldsymbol{x})\,\mathcal L_{\ell'}(\hat{\boldsymbol{p}}\,\hat{\boldsymbol{x}}).
\end{equation}
The window and the shot noise then read
\begin{align}
W_{{\rm IC},\ell\ell'} (\boldsymbol k, \boldsymbol{p}) &= W_{\ell\ell'}(\boldsymbol{k}, \boldsymbol{p}) - \sum_m v_m^{(\ell)}(\boldsymbol{k})\,e_m^{(\ell')*}(\boldsymbol{p}), \nonumber\\
    S_{{\rm IC},\ell}(\boldsymbol{k}) &= \mathcal S_\ell(\boldsymbol{k}) - \sum_m v_m^{(\ell)}(\boldsymbol{k})\,v_m^*(\boldsymbol{k}),
\end{align}
where $W_{\ell\ell'}(\boldsymbol{k}, \boldsymbol{p})$ and $\mathcal S_{\ell}(\boldsymbol{k})$ are the window and shot noise contributions in the absence of any integral constraints. The window matrix and shot noise correction are given in terms of the above in Eq.~\eqref{eq:WindowMatrixWrittenOut} and Eq.~\eqref{eq:WindowAndNoise}.

\end{document}